\documentclass[preprint,superscriptaddress,amsmath,amssymb,aps,eqsecnum,nofootinbib]{revtex4}
\usepackage[dvipdfmx]{graphicx}

\usepackage{subfigure}
\usepackage{bm}
\usepackage{mathrsfs}
\usepackage{amsmath}
\usepackage{cases}

\usepackage{color}
\usepackage{ulem}

\begin{document}
\title{Stability of relativistic tidal response against small potential modification}
\author{Takuya Katagiri}
\affiliation{Astronomical Institute, Graduate School of Science, Tohoku University, Aoba, Sendai 980-8578, Japan}
\affiliation{Niels Bohr International Academy, Niels Bohr Institute, Blegdamsvej 17, 2100 Copenhagen, Denmark}

\author{Hiroyuki Nakano}
\affiliation{Faculty of Law, Ryukoku University, Kyoto 612-8577, Japan}
\author{Kazuyuki Omukai}
\affiliation{Astronomical Institute, Graduate School of Science, Tohoku University, Aoba, Sendai 980-8578, Japan}
\date{\today}

\begin{abstract}
The tidal response of compact objects in an inspiraling binary system is measured by a set of tidal Love and dissipation numbers imprinted in the gravitational waveforms. While a four-dimensional black hole in vacuum within General Relativity has vanishing Love numbers, a black hole in alternative theories of gravity can acquire non-vanishing Love numbers. The dissipation numbers may quantify Planckian corrections at the horizon scale. These properties will allow a test of classical theories of gravity in the strong-field regime with gravitational-wave observation. Since black holes are not in the exact vacuum environment in astrophysical situations, the following question arises: can the environment affect the tidal response? In this paper, we investigate the stability of the tidal response of a Schwarzschild black hole for frequency-dependent tidal-field perturbations against a small modification of the background. Our analysis relies on the scattering theory, which overcomes difficulties in computing the relativistic tidal Love numbers. The tidal Love and dissipation numbers can be understood from the property of sufficiently low-frequency scattering waves. We show that the tidal Love numbers are sensitive to the property of the modification. Therefore, we need careful consideration of the environment around the black hole in assessing the deviation of the underlying theory of gravity from General Relativity with the Love numbers. The modification has less impact on the dissipation numbers, indicating that quantifying the existence of the event horizon with them is not spoiled. We also demonstrate that in a composite system, i.e., a compact object with environmental effects, the Love and dissipation numbers are approximately determined by the sum of the numbers of each component.

\end{abstract}
\maketitle

\tableofcontents

\section{Introduction and summary}
\label{Section:IntroductionAndSummary}
After the first detection of the binary black-hole merger~GW150914~\cite{LIGOScientific:2016aoc,LIGOScientific:2016vbw}, the advanced Laser Interferometer Gravitational-Wave Observatory~(LIGO)~\cite{LIGOScientific:2014pky} and the Virgo interferometer~\cite{VIRGO:2014yos} have already detected around a hundred coalescences of binary black holes, binary neutron stars, and binary black hole-neutron star. This observational success has given new insights into astrophysics~\cite{Coulter:2017wya,LIGOScientific:2017bnn,LIGOScientific:2018jsj}, fundamental physics~\cite{LIGOScientific:2016lio,LIGOScientific:2018dkp,Berti:2018vdi,Berti:2018cxi,Barack:2018yly}, and cosmology~\cite{LIGOScientific:2017adf,Sakstein:2017xjx}. The detection of gravitational waves from the binary neutron-star merger~GW170817~\cite{LIGOScientific:2017vwq}, which is associated with the electromagnetic counterpart, i.e., a short gamma-ray burst and a kilonova~\cite{LIGOScientific:2017zic,Kasen:2017sxr}, has opened up the field of multi-messenger astronomy~\cite{LIGOScientific:2017ync}, and provided a stringent constraint on the propagation speed of the graviton~\cite{Baker:2017hug,LIGOScientific:2017zic}. Future observations with improved detector sensitivity and/or new facilities including KAGRA~\cite{Somiya:2011np}, may provide the decisive answer to one of the most fundamental questions: to what extent is General Relativity correct in the strong-gravity regime?

In binary coalescences, the tidal interaction deforms their bodies, thereby modifying their orbital motion in the last stage of the inspiral phase. Tidal deformability at the linear level is quantified by a set of the so-called {\it tidal Love numbers} as the response of the object to an external static tidal field~\cite{Hinderer:2007mb,Damour:2009vw,Binnington:2009bb}. 
The phase evolution of gravitational waves from the inspiraling binary reflects the underlying theory of gravity as well as the internal structure of the objects through the tidal Love numbers. Measurement of them thus allows one to access the strong-field gravity of black holes or the extremely dense inside environment of neutron stars~\cite{Flanagan:2007ix,Vines:2011ud,LIGOScientific:2018cki}. 

Intriguingly, the Love numbers vanish exactly in theory for a black hole in vacuum in four-dimensional General Relativity, i.e., Schwarzschild and Kerr black holes~\cite{Binnington:2009bb,Hui:2020xxx,LeTiec:2020bos,LeTiec:2020spy,Chia:2020yla,Charalambous:2021mea}. The measurement of the tidal Love numbers of compact binaries can thus put a constraint on quantum corrections of the event horizon~\cite{Pani:2015tga,Cardoso:2017cfl,Maselli:2018fay,Addazi:2018uhd,Cardoso:2019rvt,Cardoso:2019nis,Kim:2020dif,Narikawa:2021pak,Nair:2022xfm} and work for testing theories of gravity in the strong-field regime~\cite{Cardoso:2017cfl,Cardoso:2018ptl,Chakravarti:2018vlt,Cardoso:2019vof,DeLuca:2022tkm,Brown:2022kbw}. In other words, a non-zero tidal Love number can be evidence of new physics. From the theoretical point of view, various authors~\cite{Porto:2016zng,Penna:2018gfx,Charalambous:2021kcz,Hui:2021vcv,BenAchour:2022uqo,Hui:2022vbh,Charalambous:2022rre,Katagiri:2022vyz,Berens:2022ebl,Kehagias:2022ndy} have argued a connection between the vanishing of the Love numbers and a hidden symmetric structure of the equation for linear gravitational perturbations around a black hole. 

At a practical level, the following concern still remains regarding the usefulness of the tidal Love numbers. Although the vanishing of the Love numbers of Schwarzschild and Kerr black holes is based on the exact vacuum assumption, in realistic situations, tidal interactions are more or less inevitably accompanied by the environmental effect such as the presence of surrounding matter, giving rise to non-zero Love numbers~\cite{Barausse:2014tra,Cardoso:2019upw,Cardoso:2021wlq,DeLuca:2021ite,DeLuca:2022xlz}. The measured Love numbers would exhibit strong sensitivity to the property of small environmental effects and its deviation from zero would be much larger than the scale of the environment, indicating {\it destabilization} of the tidal Love numbers. Such destabilization may interfere with the test of General Relativity.
It is thus important to theoretically assess the property of the tidal Love number induced by the deviation from the vacuum environment around a black hole within General Relativity in advance of future gravitational-wave observation.\footnote{A similar problem has been discussed in the context of quasinormal modes in Refs.~\cite{Jaramillo:2020tuu,Cheung:2021bol,Berti:2022xfj,Kyutoku:2022gbr} in the past few years and in Ref.~\cite{Nollert:1996rf} a long time ago.}  

In such a situation, a strategy within a theory/model-agnostic framework will be useful because no one knows either the ``true'' theory of gravity or the accurate model of surrounding matter fields around astrophysical black holes. However, one immediately encounters a difficulty in evaluating the relativistic tidal Love numbers, except for exactly solvable systems, because it relies on the property of the hypergeometric function: the tidal Love numbers are understood from the connection coefficient between two fundamental solutions for a tidal field, which are constructed, respectively, around the event horizon and around large distances, based on the analytic continuation of the hypergeometric or associated functions~\cite{Binnington:2009bb,Hui:2020xxx,LeTiec:2020bos,LeTiec:2020spy,Chia:2020yla,Charalambous:2021mea} as will be seen in Sec.~\ref{Section:tidalresponseofSchwarzschildBH}. Without an analytic solution capturing the physical property of a tidal field near the event horizon, it is difficult to obtain the corresponding coefficients in general, even in a system slightly modified from the Schwarzschild or Kerr background. Furthermore, even if one tries to construct local solutions analytically, for example, with the Frobenius method, there is another problem stated in the following.

One here must address a more fundamental problem: a potential ambiguity in computation of the relativistic tidal Love numbers~\cite{Damour:2009vw,Kol:2011vg,Gralla:2017djj}. To see this, let us consider the asymptotic behavior of the quadrupolar mode~$\Phi_2$ of the static gravitational perturbation to an asymptotically flat spacetime at large distances, which schematically takes the form,
\begin{equation}
\label{example:quadrupolarLove}
\Phi_2 \propto r^{3}\left[1+\mathcal{O}\left(1/r\right)\right]+\frac{\kappa}{r^2}\left[1+\mathcal{O}\left(1/r\right)\right],
\end{equation}
where $\kappa$ is the ratio of the coefficient of the decaying term in $r$ to that of the growing term, corresponding to the quadrupolar tidal Love number~\cite{Hinderer:2007mb,Damour:2009vw,Binnington:2009bb}~(see the precise definition in Sec.~\ref{Section:tidalLoveandDissipationnumbers}). However, if one considers the subleading contribution of the term of $r^3$, say, $r^3[1+\cdots+\delta_5/r^5+\mathcal{O}(1/r^6)]$, that ratio changes, i.e., $\kappa\to\kappa+\delta_5$. That is, the coefficient of the decaying term degenerates with that of the subleading term of the growing term.\footnote{The Newtonian tidal Love numbers are unambiguously computed because solutions of the Laplace equation have no such subleading terms~\cite{poisson_will_2014}. } This ambiguity leads to the discrepancy between the theoretical definition within the linear gravitational perturbation theory and the observational one appearing in gravitational-waveform models based on the post-Newtonian approximation~(see the details in Ref.~\cite{Gralla:2017djj}). 

An analytic continuation of the multipolar index from an integer to a generic number allows us to avoid the aforementioned degeneracy~\cite{Hui:2020xxx,LeTiec:2020bos,LeTiec:2020spy,Chia:2020yla,Charalambous:2021mea,Creci:2021rkz}; however, for static perturbations, another problem, i.e., the gauge-dependence of the tidal Love numbers~\cite{Gralla:2017djj}, still remains. On the other hand, as pointed out in Ref.~\cite{Creci:2021rkz}, together with the analytic continuation, the extraction of them from frequency-dependent perturbations bypasses the gauge ambiguity because of the imposition of boundary conditions that capture both the physical properties of the compact object and the perturbation field~(see also the details in Ref.~\cite{Chia:2020yla}). Despite these efforts, there still remains, in general, yet another subtlety of the definition: perturbation fields in generic systems do not necessarily take the simple form of Eq.~\eqref{example:quadrupolarLove} at large distances, instead include the logarithmic term~\cite{Kol:2011vg,Hui:2020xxx,Charalambous:2021mea,DeLuca:2022tkm}. Recently, to avoid these issues, several authors have attempted to evaluate the linear response in terms of scattering amplitude~\cite{Creci:2021rkz,Ivanov:2022qqt} and the worldline effective field theory approach~\cite{Ivanov:2022hlo}.

In this paper, we study the property of the destabilization effect on the tidal Love numbers, introduced by tiny deformation to the Schwarzschild background. To overcome the difficulties in computing the relativistic tidal Love numbers, our analysis relies on the scattering theory of linear frequency-dependent gravitational fields around a Schwarzschild black hole. We also discuss stability of another quantity characterizing frequency-dependent tidal response, i.e., the {\it dissipation number}~\cite{LeTiec:2020bos,LeTiec:2020spy,Chia:2020yla,Charalambous:2021mea,Ivanov:2022hlo}~(see the precise definition in Sec.~\ref{Section:tidalLoveandDissipationnumbers}), which quantifies the absorption of the external tidal field into the event horizon and is imprinted in gravitational waveforms from an inspiraling binary~\cite{Tagoshi:1994sm,Alvi:2001mx,Poisson:2004cw,Porto:2007qi,Chatziioannou:2012gq,Maselli:2017cmm,Cardoso:2019rvt}. Black holes have non-zero dissipation numbers, while the dissipation numbers vanish for non-rotating horizonless objects. This property may allow one to constrain Planckian corrections at the horizon scale~\cite{Addazi:2018uhd,Cardoso:2019rvt}.

In Sec.~\ref{Section:tidalLoveandDissipationnumbers}, after reviewing the relativistic tidal Love numbers of a Schwarzschild black hole for a static tidal field, we introduce the dissipation numbers. 

We first show in Sec.~\ref{Section:tidalresponseofSchwarzschildBH} that the tidal Love and dissipation numbers can be extracted from a {\it response function}~\eqref{responseF} that is defined from a low-frequency scattering wave, giving another physical interpretation for them as a property of the scattering wave~\cite{Chia:2020yla,Creci:2021rkz,Ivanov:2022qqt}. We further derive the formulas for the Love and dissipation numbers, i.e., Eqs.~\eqref{responseFandLoveNumbers} and~\eqref{responseFandDissipationNumbers}, in terms of the response function, which allows us to provide further insights into them from the viewpoint of the scattering problem. Following Refs.~\cite{Chia:2020yla,Creci:2021rkz,Ivanov:2022qqt,Ivanov:2022hlo}, our discussion overcomes the potential gauge-ambiguity in computation of the relativistic tidal Love numbers by defining them in terms of scattering waves under the boundary conditions that capture the physical property of a compact object in the strong-field regime and of a scattering wave in the weak-field regime. In addition, we use the analytic continuation of the multipole number from an integer to a generic number, which resolves the aforementioned degeneracy~\cite{Creci:2021rkz}. 

In Sec.~\ref{Section:tidalresponseofdeformedSchwarzschild}, we study the tidal response of the Schwarzschild black hole with a slight potential modification in the shape of a Gaussian small bump~(see Fig.~\ref{DeformedPotential}). The tidal Love and dissipation numbers are calculated from the response function numerically. We find that the bump leads to a non-zero Love number, while it has less impact on the dissipation number. The deviation from the vanishing Love numbers is much larger than the scale of the Gaussian bump and is quite sensitive to the property of the bump, i.e., the location, height, and width, while the dissipation number is not. This shows that the tidal Love numbers are easily destabilized by a tiny deviation from the exact vacuum environment, while the dissipation numbers are stable. 

{\it Why does a tiny modification to the Schwarzschild background destabilize the tidal Love numbers, but not the dissipation numbers?} This comes from the fact that the tidal response consists of both those of the black hole and the Gaussian bump. In a composite system, i.e., a Schwarzschild black hole with environmental effects, the tidal Love and dissipation numbers are approximately determined by the linear combination of the numbers of each component. 

As shown in Appendix~\ref{Appendix:BHplusBumpinMinkowski}, a Gaussian bump in the Minkowski spacetime causes a non-zero Love number and zero dissipation number. In fact, the former takes a value close to that of the Schwarzschild black hole with the same Gaussian bump and shares the qualitatively same feature, e.g., the dependence on the location of the bump~(see Figs.~\ref{LimitOfReResponseFunction_GaussianinMinkowski_a} and~\ref{LimitOfReResponseFunction_GaussianinMinkowski_h}). The latter has less impact on the dissipation numbers of the black hole.

Furthermore, we analyze in Sec.~\ref{Section:tidalresponseofdeformedSchwarzschild} the tidal response of a Schwarzschild black hole with a combination of the two of Gaussian bump and/or dip~(see Fig.~\ref{DeformedPotential2}). It is demonstrated in Appendix~\ref{Appendix:potentialbymatterfield} that such a combined modification can be realized due to the presence of local matter. We show that the tidal Love and dissipation numbers of the composite system are almost the same as the linear combination of the numbers for each modification. It is also found that the dissipation numbers are still stable. 

In Sec.~\ref{Section:discussion}, we discuss the astrophysical implication of our results. The destabilization indicates that a test of theories of gravity with the tidal Love numbers may require careful consideration of the environment in which black holes are immersed. From another viewpoint, a non-zero Love number allows us to catch a glimpse of the extreme property of matter fields around a black hole through gravitational-wave observation. In yet another context, the destabilization of the Love numbers may hinder constraining the matter equation of state in neutron stars because the destabilization also occurs even for horizonless compact objects such as a neutron star as seen in Appendix~\ref{Appendix:HCOcase}. For an inspiraling binary, the environmental effect varies with time as the orbital separation decreases. Consequently, the tidal response induced by the environment varies with time, while that arising from modification in theories of gravity remains constant. Thus, the extraction of the constant component from the time-varying tidal response will be an important step in testing theories of gravity in the strong-field regime. We also discuss theoretical application to ``parametrized'' formalism for the effective potential as in Refs.~\cite{Cardoso:2019mqo,McManus:2019ulj,Volkel:2022aca}.

In Appendix~\ref{Appendix:HCOcase},
we discuss the case of horizonless compact objects that have a reflective boundary at a slightly larger radius than the Schwarzschild radius.\footnote{Horizonless compact objects have been discussed as one possibility of signatures of a quantum correction in the strong-gravity regime, see Refs.~\cite{Cardoso:2016rao,Cardoso:2016oxy,Cardoso:2019apo,Agullo:2020hxe,Sago:2021iku}. } We find that a Gaussian bump destabilizes the tidal Love numbers and still maintains the vanishing of the dissipation numbers. The value of the tidal Love number of the total system is mostly determined by the Gaussian bump. Thus, it may be challenging to distinguish horizonless compact objects from black holes in the presence of environmental effects in terms of the Love numbers. On the other hand, the dissipation numbers are stable, and therefore tells their difference.

\section{Tidal response of a Schwarzschild black hole}
\label{Section:tidalLoveandDissipationnumbers}

In this section, we review linear gravitational perturbation theory of a Schwarzschild black hole, and then introduce relativistic tidal Love and dissipation numbers based on Refs.~\cite{Hinderer:2007mb,Damour:2009vw,Binnington:2009bb}. 

\subsection{Linear gravitational perturbation theory}
To discuss the tidal response of a Schwarzschild black hole to an external tidal field, we review linear gravitational perturbation theory  around the black hole.
In spherical polar coordinates~$(t,r,\theta,\varphi)$, the metric of the Schwarzschild black hole spacetime is given by
\begin{equation}
\label{Schwarzschildmetric}
g_{\mu\nu}^{(0)}dx^\mu dx^\nu=-\left(1-\frac{2M}{r}\right)dt^2+\left(1-\frac{2M}{r}\right)^{-1}dr^2+r^2\left(d\theta^2+\sin^2\theta d\varphi^2\right),
\end{equation}
where $M$ is the Arnowitt-Deser-Misner mass. Here, the coordinate range is restricted to $-\infty<t<\infty$, $2M<r<\infty$, $0\leq \theta<\pi$, and $0\leq \varphi<2\pi$. The linearly perturbed spacetime metric is given by
\begin{equation}
\label{metricgmunu}
g_{\mu\nu} =g_{\mu\nu}^{(0)}+h_{\mu\nu},
\end{equation}
where the perturbation field~$h_{\mu\nu}$ satisfies $|h_{\mu\nu}|\ll |g_{\mu\nu}^{ (0)}|$. 

Given the spherical symmetry of the background spacetime~\eqref{Schwarzschildmetric}, each component of $h_{\mu\nu}$ can be decomposed in terms of the tensor spherical harmonics. The perturbation field can be separated into even- and odd-parity modes, $h_{\mu\nu}^{\rm (even)}$ and $h_{\mu\nu}^{\rm (odd)}$, respectively, subject to the parity transformation~${\bf P}:(\theta,\varphi)\to(\pi-\theta,\varphi+\pi)$. There is no mixing between the even- and odd-parity modes because the Schwarzschild metric~\eqref{Schwarzschildmetric} is invariant under the parity transformation~${\bf P}$; one can therefore treat each parity mode independently. 

With appropriate gauge fixing, i.e., the so-called Regge-Wheeler gauge~\cite{Regge:1957td}, the harmonic modes of $h_{\mu\nu}$ in the frequency domain can be described by
\begin{equation}
\left(h_{\ell m}^{\rm (even)}\right)_{\mu\nu}=
\begin{pmatrix}
\left(1-\dfrac{2M}{r}\right)H_0^{\ell m} Y^{\ell m}& H_1^{\ell m} Y^{\ell m} & 0 &0\\
 H_1^{\ell m} Y^{\ell m}  &\left(1-\dfrac{2M}{r}\right)^{-1}H_2^{\ell m} Y^{\ell m} &0 & 0\\
0 & 0 & r^2 K^{\ell m} Y^{\ell m} &0\\
0 & 0 &0 &  r^2 \sin^2\theta K^{\ell m} Y^{\ell m}
\end{pmatrix},
\end{equation}
and
\begin{equation}
\left(h_{\ell m}^{\rm (odd)}\right)_{\mu\nu}=
\begin{pmatrix}
0& 0 & h_0^{\ell m} S_{\theta}^{\ell m} &h_0^{\ell m} S_{\varphi}^{\ell m}\\
 0 &0 & h_1^{\ell m} S_{\theta}^{\ell m} &  h_1^{\ell m} S_{\varphi}^{\ell m}\\
 h_0^{\ell m} S_{\theta}^{\ell m} &  h_1^{\ell m} S_{\theta}^{\ell m} & 0 &0\\
 h_0^{\ell m} S_{\varphi}^{\ell m} &  h_1^{\ell m} S_{\varphi}^{\ell m} &0 & 0
\end{pmatrix},
\end{equation}
where $Y^{\ell m}$ are the scalar spherical harmonics, and $(S_\theta^{\ell m},S_{\varphi}^{\ell m}):=(-\partial_\varphi Y^{\ell m}/\sin\theta, \sin\theta\partial_\theta Y^{\ell m})$. This form is valid only for $\ell\ge 2$. Here, the Fourier transformation with respect to the time variable has been performed as
\begin{equation}
{F}\left(r;\omega\right)=\int_{-\infty}^\infty dt \tilde{F}\left(t,r\right) e^{i\omega t},
\end{equation}
where ${F}$ and $\tilde{F}$ correspond to $H_0^{\ell m}, H_1^{\ell m}, H_2^{\ell m}, K^{\ell m}, h_{0}^{\ell m}$, and $h_1^{\ell m}$ in the frequency domain and those in the time domain, respectively. 

The vacuum linearized Einstein equation, $\delta R_{\mu\nu}=0$, leads to two independent equations, respectively, for the even- and odd-parity modes, which take the following unified form in the frequency domain~\cite{Regge:1957td,Zerilli:1970wzz}:
\begin{equation}
\label{RWZeqs}
\left(1-\frac{2M}{r}\right) \frac{d}{dr}\left[\left(1-\frac{2M}{r}\right)\frac{d}{dr}\Phi_{\ell m}^{\pm}\left(r;\omega\right)\right]+\left(\omega^2-V_{\ell m}^{\pm}\right)\Phi_{\ell m}^{\pm}\left(r;\omega\right)=0,
\end{equation}
with
\begin{equation}
\begin{split}
V_{\ell m}^+:=&\left(1-\frac{2M}{r}\right)\left[\frac{2\lambda^2\left(1+\lambda\right)r^3+6\lambda^2Mr^2+18\lambda M^2r+18M^3}{r^3\left(\lambda r+3M\right)^2}\right],\\
V_{\ell m}^-:=&\left(1-\frac{2M}{r}\right)\left[\frac{\ell\left(\ell+1\right)}{r^2}-\frac{6M}{r^3}\right],
\end{split}
\end{equation}
where $\lambda:=(\ell-1)(\ell+2)/2$. Here, $\Phi_{\ell m}^+$ and $\Phi_{\ell m}^-$ are gauge invariant variables for the even- and odd-parity modes, respectively, and are defined by~\cite{1974AnPhy..88..323M}
\begin{equation}
\begin{split}
\label{Phipm}
\Phi_{\ell m}^+:=&\frac{r(r-2M)}{\left(\lambda+1\right)\left(\lambda r+3M\right)}\left(H_2^{\ell m}-r\frac{dK_{\ell m}}{dr}+\frac{\lambda r+3M}{r-2M}K^{\ell m}\right),\\
\Phi_{\ell m}^-:=&\frac{r}{\lambda}\left[r^2\frac{d}{dr}\left(\frac{h_0^{\ell m}}{r^2}\right)+i\omega h_1^{\ell m}\right].
\end{split}
\end{equation}
Equations for $\Phi_{\ell m}^{+}$ and $\Phi_{\ell m}^{-}$ in Eq.~\eqref{RWZeqs} are called the Zerilli/Regge-Wheeler equations, respectively~\cite{Zerilli:1970wzz,Regge:1957td}. One can reconstruct all the nonvanishing components of $(h_{\ell m}^{\rm (even)})_{\mu\nu}$ and $(h_{\ell m}^{\rm (odd)})_{\mu\nu}$ from $\Phi_{\ell m}^+$ and $\Phi_{\ell m}^-$, respectively,
\begin{equation}
\begin{split}
H_0^{\ell m}=&\frac{1}{\lambda r+3M}\left\{\left[\frac{\lambda\left(\lambda r^2+6M^2\right)}{r\left(\lambda r+3M\right)}+\frac{3M^2}{r^2}+\lambda^2-\omega^2r^2\frac{\lambda r+3M}{r-2M}\right]\Phi_{\ell m}^+\right.\\
&\left.~~~~~~~~~~~~~-\left[\frac{M\left(\lambda r+3M\right)}{r}-\lambda \left(r-2M\right)\right]\frac{d\Phi_{\ell m}^+}{dr}\right\},\\
H_1^{\ell m}=&-i\omega\frac{\lambda r\left(r-2M\right)-M\left(\lambda r+3M\right)}{\left(r-2M\right)\left(\lambda r+3M\right)}\Phi_{\ell m}^+-i\omega r\frac{d\Phi_{\ell m}^+}{dr},\\
H_2^{\ell m}=&H_0^{\ell m},\\
K^{\ell m}=&\frac{\lambda \left(\lambda+1\right)r^2+3\lambda M r+6M^2}{r^2\left(\lambda r+3M\right)}\Phi_{\ell m}^++\left(1-\frac{2M}{r}\right)\frac{d\Phi_{\ell m}^+}{dr},
\end{split}
\end{equation}
and
\begin{equation}
\begin{split}
h_0^{\ell m}=&\left(1-\frac{2M}{r}\right)\frac{d}{dr}\left(r\Phi_{\ell m}^-\right),\\
h_1^{\ell m}=&-\frac{i\omega r^2}{r-2M}\Phi_{\ell m}^{-}.
\end{split}
\end{equation}

\subsection{Definition: tidal Love numbers}
In this subsection, the tidal Love numbers are introduced within a relativistic framework. Assuming that the tidal field is weak and slowly varying in time, we can apply linear static gravitational perturbation theory. To define the
tidal Love numbers for the static perturbation, we first introduce the notion of induced multipole moments and tidal moments of a generic metric function following Refs.~\cite{RevModPhys.52.299,Hinderer:2007mb,Cardoso:2017cfl}. 

In asymptotically Cartesian and mass centered coordinates~$(t,r,\theta,\varphi)$, induced multipole moments and tidal moments of any static, spherically symmetric, and asymptotically flat spacetime can be extracted from the asymptotic behavior of the metric components in the asymptotically flat region:
\begin{equation}
\begin{split}
\label{gttgtvarphi}
g_{tt}=&-1+\frac{2M}{r}-\sum_{\ell\ge 2}\left[\frac{2}{\ell\left(\ell-1\right)}r^\ell\left(\mathcal{E}_{\ell}Y^{\ell0}+\left(\ell>\ell'\right)\right) \right.
\\ & \qquad \qquad \qquad \qquad
\left.
-\frac{2}{r^{\ell+1}}\left(\sqrt{\frac{4\pi}{2\ell+1}}{\cal M}_{\ell}Y^{\ell 0}+\left(\ell>\ell'\right)\right)\right],\\
g_{t\varphi}=&\sum_{\ell\ge 2}\left[\frac{2}{3\ell\left(\ell-1\right)}r^{\ell+1}\left(\mathcal{B}_\ell S_\varphi^{\ell 0}+\left(\ell>\ell'\right)\right)+\frac{2}{r^\ell}\left(\sqrt{\frac{4\pi}{2\ell+1}}\frac{{\cal S}_\ell}{\ell}S_{\varphi}^{\ell0}+\left(\ell>\ell'\right)\right)\right],
\end{split}
\end{equation}
where we have defined $M$ as the Arnowitt-Deser-Misner mass of the central gravitational source; ${\cal E}_\ell$ and ${\cal B}_\ell$ as even- and odd-parity components of tidal moments, respectively, which correspond to the amplitudes of even- and odd-parity modes of the external tidal field, respectively; ${\cal M}_\ell$ and ${\cal S}_\ell$ as the induced mass multipole moments and induced current multipole moments, respectively. The notation of $\left(\ell>\ell'\right)$ denotes the contribution of $\ell'$ ($<\ell$) poles.

We now define components of the {\it tidal Love numbers} for even- and odd-parity modes, respectively, as~\cite{Cardoso:2017cfl}
\begin{equation}
\begin{split}
\label{kappaEkappaB}
\kappa_\ell^{E}:=&-\frac{\ell\left(\ell-1\right)}{2r_0^{2\ell+1}}\sqrt{\frac{4\pi}{2\ell+1}}\frac{{\cal M}_\ell}{{\cal E}_\ell},\\
\kappa_\ell^{B}:=&-\frac{3\ell\left(\ell-1\right)}{2\left(\ell+1\right)r_0^{2\ell+1}}\sqrt{\frac{4\pi}{2\ell+1}}\frac{{\cal S}_\ell}{{\cal B}_\ell},
\end{split}
\end{equation}
where $r_0$ is the radius of the central gravitational source. These are also called {\it electric-type} and {\it magnetic-type tidal Love numbers}, respectively. The tidal Love numbers correspond to the dimensionless ratio of the coefficient of the growing part in $r$ to that of the decaying part of the asymptotic behavior of the metric components in Eq.~\eqref{gttgtvarphi}. 

For the Schwarzschild black hole, the tidal Love numbers can be read off from $\Phi_{\ell m}^{\pm}$~\cite{Hui:2020xxx}, 
\begin{equation}
\begin{split}
\label{LoveNumbersinZRWvariables}
\left.\Phi_{\ell m}^\pm\right|_{\omega=0, r\to\infty}&\sim \left(\frac{r}{2M}\right)^{\ell+1}\biggl\{1+\mathcal{O}\left(2M/r\right)
\\ & \qquad \qquad \qquad \quad
+2\frac{\left(\ell+2\right)\left(\ell+1\right)}{\ell\left(\ell-1\right)}\kappa_{\ell}^{E/B}\left(\frac{r}{2M}\right)^{-2\ell-1}\left[1+\mathcal{O}\left(2M/r\right)\right]\biggr\}.
\end{split}
\end{equation}
Note that $\Phi_{\ell m}^{\pm}$ are required to be regular at the black hole horizon~$r=2M$. The well-known intriguing result is that the horizon-regular solutions of $\Phi_{\ell m}^{\pm}$ have no decaying series in $r$, i.e., $\Phi_{\ell m}^{\pm}|_{r\gg 2M}\propto [r/(2M)]^{\ell+1}[1+{\cal O}(2M/r)]$, indicating the vanishing of the tidal Love numbers, i.e., $\kappa_\ell^{E}=\kappa_\ell^{B}=0$~\cite{Hinderer:2007mb,Damour:2009vw,Binnington:2009bb}.

\subsection{Definition: dissipation numbers}
We here also introduce the relativistic dissipation numbers. Now, an external tidal field is assumed to be weak and has a low frequency~$\omega\ll 1/(2M)$. The dissipation numbers of a Schwarzschild black hole can then be read off from $\Phi^{\pm}_{\ell m}$ at large distances as follows. Under the requirement of no outgoing tidal field from the event horizon, the variables~$\Phi^{\pm}_{\ell m}$ with a low frequency~$\omega\ll1/(2M)$ at large distances~$2M\ll r \ll1/\omega$ take the following forms (see Sec.~\ref{Subsection:tidalLovefromphaseshiftinanalytic}):
\begin{equation}
\begin{split}
\label{LoveandDissipationNumbersinZRWvariables}
\left.\Phi_{\ell m}^\pm \right|_{2M\ll r\ll 1/\omega}&\sim \left(\frac{r}{2M}\right)^{\ell+1}\biggl\{1+\mathcal{O}\left(2M/r\right)
\\ & \qquad \qquad \qquad \quad
+2\frac{\left(\ell+2\right)\left(\ell+1\right)}{\ell\left(\ell-1\right)}F_\ell^{E/B}\left(\omega\right)\left(\frac{r}{2M}\right)^{-2\ell-1}\left[1+\mathcal{O}\left(2M/r\right)\right]\biggr\},
\end{split}
\end{equation}
with the functions~$F_\ell^{E/B}$ of $\omega$ defined by
\begin{equation}
\label{ExpansionOfKfunctions}
F_\ell^{E/B}=\kappa_\ell^{E/B}+\frac{i}{2}\nu_\ell^{E/B} (2\omega M) +{\cal O}\left((2\omega M)^2\right).
\end{equation}
Here, $\kappa_\ell^{E/B}$ are the electric-type/magnetic-type tidal Love numbers of the Schwarzschild black hole for each $\ell$-pole mode, and then $\kappa_\ell^{E}=\kappa_\ell^{B}=0$~\cite{Hinderer:2007mb,Damour:2009vw,Binnington:2009bb}. 

The quantities~$\nu_\ell^{E/B}$ correspond to {\it electric-type/magnetic-type dissipation numbers}, which quantify the dissipation of the tidal field into the event horizon. Schwarzschild black holes have non-zero dissipation numbers owing to the presence of the event horizon~\cite{LeTiec:2020bos,LeTiec:2020spy,Chia:2020yla,Charalambous:2021mea,Ivanov:2022hlo}. This is also the case for Kerr black holes~\cite{Chia:2020yla,Charalambous:2021mea,Ivanov:2022hlo}.\footnote{Kerr black holes have non-zero dissipation numbers even for static perturbations because of a relative motion with a static environment, which is sourced by rotation~\cite{Chia:2020yla,Charalambous:2021mea,Ivanov:2022hlo}.}  On the other hand, the dissipation numbers of a non-rotating horizonless object vanish due to the absence of the event horizon. Those properties allow one to quantify the existence of the event horizon in the context of testing classical theories of gravity in the strong-field regime~\cite{Pani:2015tga,Cardoso:2017cfl,Maselli:2018fay,Addazi:2018uhd,Cardoso:2019rvt,Kim:2020dif,Narikawa:2021pak}.

\section{Tidal response of a Schwarzschild black hole in scattering theory}
\label{Section:tidalresponseofSchwarzschildBH}

In astrophysical applications such as to binary systems, the gravitational field generated by objects varies with time. A time-varying weak tidal field, i.e., a gravitational wave, around a Schwarzschild black hole is described by a solution of the Zerilli/Regge-Wheeler equations~\eqref{RWZeqs}. In this section, we discuss the frequency-dependent tidal response in terms of the scattering theory of gravitational waves around a Schwarzschild black hole. In particular, we show that the tidal Love and dissipation numbers of a Schwarzschild black hole can be extracted from a response function that is defined from the property of scattering waves.  Although we here focus only on the odd-parity mode, we expect to be able to extract tidal responses from scattering waves even for the even-parity mode thanks to the existence of isospectrality~\cite{Chandrasekhar:1975zza}. We have also checked that the results in this section are valid for both scalar and vector-field perturbations.

\subsection{Scattering waves around a Schwarzschild black hole}
We here introduce basics of scattering gravitational waves around a Schwarzschild black hole by focusing on the Regge-Wheeler equation in the frequency domain,
\begin{equation}
\label{RWeq}
\left(1-\frac{r_H}{r}\right)\frac{d}{dr}\left[\left(1-\frac{r_H}{r}\right)\frac{d\Phi}{dr}\right]+\left\{\omega^2-\left(1-\frac{r_H}{r}\right)\left[\frac{\ell(\ell+1)}{r^2}-\frac{3r_H}{r^3}\right]\right\}\Phi=0,
\end{equation}
where $r_H:=2M$.
Note that this is identical to one of Eq.~\eqref{RWZeqs}. Now, let us consider the situation where a monochromatic plane wave with a frequency~$\omega$ propagates along the $x$-axis toward a Schwarzschild black hole and is scattered off by its effective potential.

Analogous to the problem in quantum mechanics, the scattering wave~$\Phi$ at large distances schematically takes the form of a superposition of the incident plane wave along the $x$-axis and the outgoing spherical wave:\footnote{One can set $m=0$ in the spherical harmonics~$Y_{\ell m}$ without loss of generality because of the spherical symmetry of the background.}
\begin{equation}
\label{ScatteringWaveBC}
\sum_{\ell=2}^\infty\frac{\Phi}{r}Y_{\ell 0}\left(\theta\right) \sim e^{-i\omega x}+f(\theta)\frac{e^{i\omega r_*}}{r_*} ,
\end{equation}
where
$r_*$ is the tortoise coordinate defined as
\begin{equation}
\label{tortoiser}
r_*:=r+r_H\ln\left(\frac{r}{r_H}-1\right),
\end{equation}
and $\theta$ is the angle measured with respect to the $x$-axis. Here, the coefficient~$f(\theta)$ is called a {\it scattering amplitude}, which can be expanded in terms of the Legendre polynomial:
\begin{equation}
\label{scatteringamplitude}
f\left(\theta\right)=\sum_{\ell=2}^{\infty}\left(2\ell+1\right)\frac{e^{2i\delta_\ell\left(\omega\right)}-1}{2i\omega}P_\ell\left(\cos \theta\right).
\end{equation}
Here, $\delta_\ell$ is a complex-valued {\it phase shift}. By performing the partial-wave decomposition of the incident plane wave, the asymptotic expression for each $\ell$-pole mode of the scattering wave in Eq.~\eqref{ScatteringWaveBC} can be written as a superposition of ingoing and outgoing spherical waves:
\begin{equation}
\frac{\Phi}{r}\sim 
\frac{2\ell+1}{2i\omega r_* }\left[e^{2i\delta_\ell}e^{i\omega r_*}-\left(-1\right)^\ell e^{-i\omega r_*}\right] .
\end{equation}
The quantity~$e^{2i\delta_\ell}$ is called an {\it S-matrix} and corresponds to the ratio of the amplitude of the outgoing spherical wave to that of the ingoing spherical wave. Therefore, the S-matrix, and, in particular, the phase shift~$\delta_\ell$, have information on ``strength'' of the scattering. 

\subsection{Tidal Love and dissipation numbers from scattering waves: analytical results}
\label{Subsection:tidalLovefromphaseshiftinanalytic}
We here show analytically how tidal Love and dissipation numbers are imprinted in scattering waves. For later convenience, by introducing new parameters,
\begin{equation}
\bar{\omega}:=\omega r_H,~~z:=\omega r,
\end{equation}
we rewrite the Regge-Wheeler equation~\eqref{RWeq} as
\begin{equation}
\label{RWeqinz}
\left(1-\frac{\bar{\omega}}{z}\right)\frac{d}{dz}\left[\left(1-\frac{\bar{\omega}}{z}\right)\frac{d\Phi}{dz}\right]+\left\{1-\left(1-\frac{\bar{\omega}}{z}\right)\left[\frac{\ell(\ell+1)}{z^2}-\frac{3\bar{\omega}}{z^3}\right]\right\}\Phi=0.
\end{equation}
In the following, we solve Eq.~\eqref{RWeqinz} with the matched asymptotic expansion method~\cite{Bender_Orszag_1999}. This strategy utilizes the analytical property of the hypergeometric functions~\cite{NIST:DLMF}, in which an analytic continuation of $\ell$ from an integer to generic numbers plays an important role. Henceforth, we assume $\ell$ to be a generic number. 

First, we assume that a tidal field has low frequencies, i.e., $\bar{\omega}\ll1$. For the inspiral of a black-hole binary system, this assumption can be justified except for a short pre-merger phase. For such a low-frequency field, one can divide the exterior of the event horizon into two regions, i.e., the near $(\bar{\omega}<z\ll1)$ and far $(z\gg\bar{\omega})$ regions. We then solve Eq.~\eqref{RWeqinz} and obtain an approximate solution in each region. Finally, by matching the solutions near and far regions in an intermediate region~$(\bar{\omega}\ll z\ll1)$, we construct the global analytic solution for the scattering wave approximately.

\subsubsection{Near-region solution}
We here derive an approximate solution of Eq.~\eqref{RWeqinz} in the near region~$(\bar{\omega}<z\ll1)$. By introducing an auxiliary function,
\begin{equation}
\label{XN}
X_{\rm N}\left(z\right):=\left(1-\frac{\bar{\omega}}{z}\right)^{-i\bar{\omega}}\left(\frac{\bar{\omega}}{z}\right)^{-\ell}\Phi,
\end{equation}
and a new coordinate variable,
\begin{equation}
x:=1-\frac{\bar{\omega}}{z},
\end{equation}
the Regge-Wheeler equation~\eqref{RWeqinz} can be reduced to the following equation for $X_{\rm N}$:
\begin{equation}
\label{EqforXN:nearregion}
x\left(1-x\right)\frac{d^2X_{\rm N}}{dx^2}+\left[\gamma_{\rm N}-\left(\alpha_{\rm N}+\beta_{\rm N}+1\right)x\right]\frac{dX_{\rm N}}{dx}-\left[\alpha_{\rm N}\beta_{\rm N}+\frac{x^2-3x+3}{\left(x-1\right)^{3}}\bar{\omega}^2\right]X_{\rm N}=0,
\end{equation}
where 
\begin{equation}
\begin{split}
\alpha_{\rm N}:=&\ell+3+i\bar{\omega}+\mathcal{O}\left(\bar{\omega}^2\right),\\
\beta_{\rm N}:=&\ell-1+i\bar{\omega}+\mathcal{O}\left(\bar{\omega}^2\right),\\
\gamma_{\rm N}:=&1+2i\bar{\omega}.
\end{split}
\end{equation}
For the range of $z\ll \bar{\omega}^{1/3}$, in the last large bracket on the left-hand side of Eq.~\eqref{EqforXN:nearregion}, the second term is much smaller than the first term $\alpha_{\rm N}\beta_{\rm N}$. By neglecting this term, the general solution of Eq.~\eqref{EqforXN:nearregion} can be written in terms of the Gaussian hypergeometric function around $x=0$~\cite{NIST:DLMF}: 
\begin{equation}
X_{\rm N}=A_{\rm in, N}x^{-2i\bar{\omega}}~_2F_1\left(\alpha_{\rm N}-\gamma_{\rm N}+1,\beta_{\rm N}-\gamma_{\rm N}+1;2-\gamma_{\rm N};x\right)+A_{\rm out,N}~_2F_1\left(\alpha_{\rm N},\beta_{\rm N};\gamma_{\rm N};x\right),
\end{equation}
where $A_{\rm in,N}$ and $A_{\rm out,N}$ are functions of $\bar{\omega}$. 

Reconstructing the original variable~$\Phi$ with Eq.~\eqref{XN}, the general solution of Eq.~\eqref{RWeqinz} in the near region~$(\bar{\omega}<z\ll 1)$ is obtained as
\begin{equation}
\begin{split}
\label{GeneralPhiN}
\left.\Phi_{\rm N}\right|_{\bar{\omega}<z\ll1}=&A_{\rm in, N}\left(1-\frac{\bar{\omega}}{z}\right)^{-i\bar{\omega}}\left(\frac{\bar{\omega}}{z}\right)^\ell~_2F_1\left(\alpha_{\rm N}-\gamma_{\rm N}+1,\beta_{\rm N}-\gamma_{\rm N}+1;2-\gamma_{\rm N};1-\bar{\omega}/z\right)\\
&+A_{\rm out,N}\left(1-\frac{\bar{\omega}}{z}\right)^{i\bar{\omega}}\left(\frac{\bar{\omega}}{z}\right)^\ell~_2F_1\left(\alpha_{\rm N},\beta_{\rm N};\gamma_{\rm N};1-\bar{\omega}/z\right).
\end{split}
\end{equation}
The first and second terms correspond to the ingoing and outgoing waves at the horizon, respectively. We impose the ingoing-wave boundary condition~$A_{\rm out,N}=0$ at the horizon:
\begin{equation}
\begin{split}
\label{Nearsol}
\left.\Phi_{\rm N}\right|_{\bar{\omega}<z\ll1}=&A_{\rm in, N}\left(1-\frac{\bar{\omega}}{z}\right)^{-i\bar{\omega}}\left(\frac{\bar{\omega}}{z}\right)^\ell~_2F_1\left(\alpha_{\rm N}-\gamma_{\rm N}+1,\beta_{\rm N}-\gamma_{\rm N}+1;2-\gamma_{\rm N};1-\bar{\omega}/z\right).
\end{split}
\end{equation}

Following the conventional manner~\cite{Binnington:2009bb,Cardoso:2017cfl,Hui:2020xxx,Chia:2020yla,Charalambous:2021mea}, it can be seen that the tidal Love numbers of the Schwarzschild black hole exactly vanish, while the dissipation numbers are not zero. To see this, we investigate the asymptotic behavior of $\Phi_{\rm N}$ in Eq.~\eqref{Nearsol} at large distances~$z\gg\bar{\omega}$. For the coordinate domain of the near region~$z\ll1$, the near-region solution~\eqref{Nearsol} at large distances takes the form,
\begin{equation}
\begin{split}
\label{asymptoticPhiN}
\left.\Phi_{\rm N}\right|_{\bar{\omega}\ll z\ll1}=&A_{\rm in,N}\frac{\Gamma\left(2-\gamma_{\rm N}\right)\Gamma\left(\alpha_{\rm N}+\beta_{\rm N}-\gamma_{\rm N}\right)}{\Gamma\left(\alpha_{\rm N}-\gamma_{\rm N}+1\right)\Gamma\left(\beta_{\rm N}-\gamma_{\rm N}+1\right)}\left(\frac{z}{\bar{\omega}}\right)^{\ell+1}\left[1+i\bar{\omega}\ln z+\mathcal{O}(\bar{\omega}/z,\bar{\omega}^2)\right]\\
&\times \left\{1+\mathcal{O}\left(\bar{\omega}/z\right)+\mathcal{K}_\ell\left(\bar{\omega}\right)\left(\frac{z}{\bar{\omega}}\right)^{-2\ell-1}\left[1+\mathcal{O}\left(\bar{\omega}/z\right)\right]\right\},
\end{split}
\end{equation}
where the coefficient of the decaying term is defined as
\begin{equation}
\label{responseK}
{\cal K}_\ell\left(\bar{\omega}\right):=\frac{\Gamma\left(-2\ell-1\right)\Gamma\left(\ell+3-i\bar{\omega}\right)\Gamma\left(\ell-1-i\bar{\omega}\right)}{\Gamma\left(2\ell+1\right)\Gamma\left(-\ell+2-i\bar{\omega}\right)\Gamma\left(-\ell-2-i\bar{\omega}\right)}.
\end{equation}
Here, we have used the relations of the hypergeometric functions, Eqs.~\eqref{Formula1for2F1} and~\eqref{Formula2for2F1}. 

Comparison of Eq.~\eqref{asymptoticPhiN} with Eq.~\eqref{LoveandDissipationNumbersinZRWvariables} tells that the real part of the function~${\cal K}_\ell$ is related to the ($\ell$th-pole) magnetic-type tidal Love number in the static limit~$\bar{\omega}\to0$. In fact, one can analytically show that the function~${\rm Re}[{\cal K}_\ell]$ indeed vanishes when~$\bar{\omega}\to0$:
\begin{equation}
\begin{split}
\label{vanishingkappa}
\lim_{\bar{\omega}\to0}{\rm Re}\left[{\cal K}_\ell\right]&=\frac{\left(-1\right)^{\ell}\Gamma\left(\ell-1\right)^2\Gamma\left(\ell+3\right)}{2\Gamma\left(2\ell+1\right)\Gamma\left(2\ell+2\right)\Gamma\left(-\ell-2\right)}=0,
\end{split}
\end{equation}
due to the presence of $1/\Gamma(-\ell-2)$ from the relation $1/\Gamma(-n)=0$ for $n=0,1,2,\cdots$. We have here used the relation of the gamma functions, Eq.~\eqref{FormulaforGamma1}. The imaginary part, ${\rm Im}[{\cal K}_\ell]$, also vanishes in the limit $\bar{\omega}\to0$.
Furthermore, comparing Eq.~\eqref{asymptoticPhiN} with Eq.~\eqref{LoveandDissipationNumbersinZRWvariables}, we can see that the ($\ell$th-pole) magnetic-type dissipation number is related to the quantity~${\rm Im}[{\cal K}_\ell]/\bar{\omega}$ at~$\bar{\omega}\to0$:  
\begin{equation}
\label{ImKell}
\lim_{\bar{\omega}\to0}\frac{{\rm Im}\left[{\cal K}_\ell\right]}{\bar{\omega}}=\frac{\Gamma\left(\ell-1\right)^2\Gamma\left(\ell+3\right)^2}{\Gamma\left(2\ell+1\right)\Gamma\left(2\ell+2\right)},
\end{equation}
which is a positive value, for example, $0.20$ for the quadrupolar ($\ell=2$) mode, giving $\nu_2^B=0.0333$.

\subsubsection{Far-region solution}
We here derive an approximate solution of Eq.~\eqref{RWeqinz} in the far region~$(z\gg\bar{\omega})$. Introducing an auxiliary function,
\begin{equation}
Y_{\rm F}(z):=e^{-iz}\left(\frac{z}{\bar{\omega}}\right)^{-\ell-1}\Phi,
\end{equation}
and a coordinate variable,
\begin{equation}
y:=-2iz,
\end{equation}
Eq.~\eqref{RWeqinz} leads to
\begin{equation}
\label{EqforYF:farregion}
y\frac{d^2Y_{\rm F}}{dy^2}+(\beta_{\rm F}-y)\left(1+\epsilon_1(z)\right)\frac{d Y_{\rm F}}{dy}-\alpha_{\rm F}\left(1+\epsilon_2(z)\right)Y_{\rm F}=0,
\end{equation}
with 
\begin{align}
\alpha_{\rm F}:=&\ell+1-i\bar{\omega},\\
\beta_{\rm F}:=&2\ell+2.
\end{align}
Here, we have defined
\begin{equation}
\begin{split}
\epsilon_1(z):=&\frac{\bar{\omega}}{2\left(\ell+1+iz\right)\left(z-\bar{\omega}\right)},\\
\epsilon_2(z):=&i\bar{\omega}\frac{z\left[\ell^2-4+\bar{\omega}\left(i+2\bar{\omega}\right)\right]-\left(\ell^2-4\right)\bar{\omega}-z^2\left(i+3\bar{\omega}\right)}{2\left(\ell+1-i\bar{\omega}\right)z\left(z-\bar{\omega}\right)^2}.
\end{split}
\end{equation}
The absolute values of $\epsilon_1(z)$ and $\epsilon_2(z)$ are both much smaller than unity in the far region~($z\gg\bar{\omega}$) for $|\bar{\omega}(\ell^2-4)/(2(\ell+1))|\ll z^2$. By neglecting $\epsilon_1(z)$ and $\epsilon_2(z)$, Eq.~\eqref{EqforYF:farregion} takes the form of the differential equation for the confluent hypergeometric functions. The general solution of Eq.~\eqref{EqforYF:farregion} can be well approximated by a linear combination of two independent confluent hypergeometric functions around $z=0$~\cite{NIST:DLMF}: 
\begin{equation}
Y_{\rm F}=A_{+} M\left(\ell+1-i\bar{\omega},2\ell+2,-2iz\right)+A_{-} U\left(\ell+1-i\bar{\omega},2\ell+2,-2iz\right),
\end{equation}
where $M(~,~,-2iz)$ and $U(~,~,-2iz)$ are confluent hypergeometric functions, which are Kummer's and Tricomi's functions, respectively; $A_{+}$ and $A_{-}$ are functions of $\bar{\omega}$.

Reconstructing the original variable~$\Phi$, we obtain the general solution of Eq.~\eqref{RWeqinz} in the far region~$(z\gg\bar{\omega})$:
\begin{equation}
\begin{split}
\label{Farregionsol}
\left.\Phi_{\rm F}\right|_{z\gg\bar{\omega}}=&A_{+}\left(\frac{z}{\bar{\omega}}\right)^{\ell+1}e^{iz}M\left(\ell+1-i\bar{\omega},2\ell+2,-2iz\right)\\
&+A_{-} \left(\frac{z}{\bar{\omega}}\right)^{\ell+1}e^{iz}U\left(\ell+1-i\bar{\omega},2\ell+2,-2iz\right).
\end{split}
\end{equation}
Using the relations of the confluent hypergeometric functions, Eqs.~\eqref{Formula1forM} and~\eqref{Formula1forU}, the asymptotic behavior at infinity~$z\to\infty$ takes the following form:
\begin{equation}
\begin{split}
\label{Farsol}
\left.\Phi_{\rm F}\right|_{z\gg \bar{\omega}}=&A_{+}\frac{(-1)^{\ell+1}}{(-2i)^{\ell+1+i\bar{\omega}}\bar{\omega}^{\ell+1}}\frac{\Gamma\left(2\ell+2\right)}{\Gamma\left(\ell+1-i\bar{\omega}\right)}\\
&\times \left\{e^{2i\delta_\ell}e^{i\left(z+\bar{\omega}\ln \left(z/\bar{\omega}-1\right)\right)}\left[1+\mathcal{O}\left(1/z\right)\right]-\left(-1\right)^\ell e^{-i\left(z+\bar{\omega}\ln\left(z/\bar{\omega}-1\right)\right)}\left[1+\mathcal{O}\left(1/z\right)\right]\right\},
\end{split}
\end{equation}
where $\delta_\ell(\bar{\omega})$ is the complex-valued phase shift defined by
\begin{equation}
\begin{split}
\label{PhaseShift}
e^{2i\delta_\ell}=&\frac{\Gamma\left(\ell+1-i\bar{\omega}\right)}{\Gamma\left(\ell+1+i\bar{\omega}\right)}\left[1+2i\bar{\omega}\ln \left(2\bar{\omega}\right)+\mathcal{O}\left(\bar{\omega}^2\right)\right]\\
&+(-1)^{\ell+1}\frac{\Gamma\left(\ell+1-i\bar{\omega}\right)}{\Gamma\left(2\ell+2\right)}\frac{A_{-}}{A_{+}}\left\{1+\bar{\omega}\left[\pi+2i\ln \left(2\bar{\omega}\right) \right]+\mathcal{O}\left(\bar{\omega}^2\right)\right\}.
\end{split}
\end{equation}
Equation~\eqref{Farsol} implies that the leading behavior of the far-region solution~\eqref{Farregionsol} at infinity consists of a superposition of outgoing and ingoing spherical waves, i.e., 
\begin{equation}
\left.\Phi_{\rm F}\right|_{z\gg \bar{\omega}}\propto e^{2i\delta_\ell}e^{iz_*}-\left(-1\right)^\ell e^{-iz_*},
\end{equation}
where $z_*:=z+\bar{\omega}\ln (z/\bar{\omega}-1)$ is the asymptotic expression for the tortoise coordinate~\eqref{tortoiser} normalized by $\omega$.  

To match the far-region and near-region solutions, we investigate the asymptotic behavior of $\Phi_{\rm F}$ in Eq.~\eqref{Farregionsol} in the intermediate region~$(\bar{\omega}\ll z\ll 1)$, which is given by 
\begin{equation}
\begin{split}
\label{asymptoticPhiF}
\left.\Phi_{\rm F}\right|_{\bar{\omega}\ll z\ll1}
=&A_{+}\left(\frac{z}{\bar{\omega}}\right)^{\ell+1}\left[1+\mathcal{O}\left(\bar{\omega}/z\right)\right]\left\{1+\mathcal{O}\left(z\right)+{\cal F}_\ell\left(\bar{\omega}\right)\left(\frac{z}{\bar{\omega}}\right)^{-2\ell-1}\left[1+\mathcal{O}(z)\right]\right\}
\end{split}.
\end{equation}
Here, a {\it response function} is introduced as~\cite{Chakrabarti:2013xza,Chia:2020yla,Creci:2021rkz},
\begin{equation}
\label{responseF}
{\cal F}_\ell:=i\frac{\left(-1\right)^\ell}{2^{2\ell+1}\bar{\omega}^{2\ell+1}}\frac{\Gamma\left(2\ell+1\right)}{\Gamma\left(\ell+1-i\bar{\omega}\right)} \frac{A_{-}}{A_{+}}.
\end{equation}
In Eq.~\eqref{PhaseShift}, writing $A_-/A_+$ in terms of ${\cal F}_\ell$, we obtain
\begin{equation}
\begin{split}
\label{PhaseShiftandF}
e^{2i\delta_\ell}=&\frac{\Gamma\left(\ell+1-i\bar{\omega}\right)}{\Gamma\left(\ell+1+i\bar{\omega}\right)}\left[1+2i\bar{\omega}\ln \left(2\bar{\omega}\right)+\mathcal{O}\left(\bar{\omega}^2\right)\right]\\
&+i\frac{2^{2\ell+1}\bar{\omega}^{2\ell+1}\Gamma\left(\ell+1-i\bar{\omega}\right)^2}{\Gamma\left(2\ell+1\right)\Gamma\left(2\ell+2\right)}{\cal F}_\ell\left\{1+\bar{\omega}\left[\pi+2i\ln \left(2\bar{\omega}\right) \right]+\mathcal{O}\left(\bar{\omega}^2\right)\right\}.
\end{split}
\end{equation}
We have here used the relation of the gamma function, Eq.~\eqref{FormulaforGamma2}. 

Equation~\eqref{asymptoticPhiF} shows that the asymptotic behavior of the far-region solution~$\Phi_{\rm F}$ in the intermediate region~$(\bar{\omega}\ll z\ll1)$ consists of a linear combination of the growing and decaying terms with increasing $z$. The response function~${\cal F}_\ell$ corresponds to the ratio of their coefficients. Comparison of Eq.~\eqref{asymptoticPhiF} with Eq.~\eqref{LoveandDissipationNumbersinZRWvariables} suggests that the response function~${\cal F}_\ell$ captures the tidal response. In the following, we see that this is indeed the case. 

\subsubsection{Matching of near- and far-region solutions and a tidal response from the response function}
\label{subsubsec:TLNfromPhaseShift}
The global solution is constructed by matching the near-region solution~\eqref{Nearsol} with the far-region solution~\eqref{Farsol}.
We then show that the function~${\cal K}_\ell$ in Eq.~\eqref{responseK} can be extracted from the response function~${\cal F}_\ell$ in Eq.~\eqref{responseF}. We further derive the formulas for the Love and dissipation numbers in terms of the response function. 

From the asymptotic behaviors of the near-region~\eqref{asymptoticPhiN} and far-region~\eqref{asymptoticPhiF} solutions, for successful matching of the two solutions it is required that the response function~${\cal F}_\ell$ in Eq.~\eqref{responseF} coincides with the function~${\cal K}_\ell$ in Eq.~\eqref{responseK} up to the overall factor, i.e., 
\begin{equation}
\label{MatchingCondition}
{\cal F}_\ell={\cal K}_\ell.
\end{equation}
This means that the response function~${\cal F}_\ell$ captures the magnetic-type tidal Love and dissipation numbers through ${\cal K}_\ell$. 
We thus obtain simple formulas for the tidal Love and dissipation numbers in terms of ${\cal F}_\ell$:
\begin{equation}
\label{responseFandLoveNumbers}
\kappa_\ell^B=\frac{\ell\left(\ell-1\right)}{2\left(\ell+2\right)\left(\ell+1\right)}\lim_{\bar{\omega}\to0}{\rm Re}\left[{\cal F}_\ell\right],
\end{equation}
and
\begin{equation}
\label{responseFandDissipationNumbers}
\nu_\ell^B=\frac{\ell\left(\ell-1\right)}{\left(\ell+2\right)\left(\ell+1\right)}\lim_{\bar{\omega}\to0}\frac{{\rm Im}\left[{\cal F}_\ell\right]}{\bar{\omega}}.
\end{equation}
These formulas are useful for several extensions to systems that are slightly modified from the Schwarzschild spacetime due to the presence of some matter fields or modification of theories of gravity. If the modification is only in the vicinity of a black hole, the far region is less affected. For this reason, one can still use the same response function~\eqref{responseF} and can safely compute the Love and dissipation numbers from Eqs.~\eqref{responseFandLoveNumbers} and~\eqref{responseFandDissipationNumbers} without the ambiguities once we obtain the values of $A_+$ and $A_-$ in Eq.~\eqref{Farregionsol} by numerical approach, e.g., by fitting the far-region solution~\eqref{Farregionsol} with the numerical solution.\footnote{In Refs.~\cite{Creci:2021rkz,Ivanov:2022qqtX}, the computation of scattering-wave amplitudes of Schwarzschild spacetimes in terms of scalar waves has been conducted, and has then been connected to gauge-invariant quantities within an effective field theory framework, thereby obtaining the Love and dissipation numbers in a gauge-invariant manner.}

With the matching condition~\eqref{MatchingCondition}, 
Eq.~\eqref{PhaseShiftandF} yields the analytic expression for the phase shift in terms of ${\cal K}_\ell$:
\begin{equation}
\begin{split}
\label{PhaseShiftandK}
e^{2i\delta_\ell}=&\frac{\Gamma\left(\ell+1-i\bar{\omega}\right)}{\Gamma\left(\ell+1+i\bar{\omega}\right)}\left[1+2i\bar{\omega}\ln \left(2\bar{\omega}\right)+\mathcal{O}\left(\bar{\omega}^2\right)\right]\\
&+i\frac{2^{2\ell+1}\bar{\omega}^{2\ell+1}\Gamma\left(\ell+1-i\bar{\omega}\right)^2}{\Gamma\left(2\ell+1\right)\Gamma\left(2\ell+2\right)}{\cal K}_\ell\left\{1+\bar{\omega}\left[\pi+2i\ln \left(2\bar{\omega}\right) \right]+\mathcal{O}\left(\bar{\omega}^2\right)\right\},
\end{split}
\end{equation}
This shows that tidal Love and dissipation numbers are imprinted in the phase shift in principle. However, the Love and dissipation numbers are at least at ${\cal O}(\bar{\omega}^{2\ell+1})$ and ${\cal O}(\bar{\omega}^{2\ell+2})$, respectively, indicating they degenerate with the subleading corrections in the first line on the right-hand side in Eq.~\eqref{PhaseShiftandK}. This means that the extraction of a tidal response from the phase shift is technically challenging.

\subsection{Tidal Love and dissipation numbers from scattering waves: numerical results}
We here numerically show that the matching condition~\eqref{MatchingCondition} is indeed satisfied, and then demonstrate that ${\rm Re}[{\cal F}_\ell]$ and ${\rm Im}[{\cal F}_\ell]/(\omega r_H)$ go to, respectively, zero and the value given in Eq.~\eqref{ImKell} in the limit~$\bar{\omega}\to0$ by extrapolating them from a non-zero frequency. 

In the numerical analysis, we use {\it Mathematica}, and integrate the Regge-Wheeler equation~\eqref{RWeq} from a radius slightly outside of the event horizon, $r=2M(1+10^{-5})$, to large distances~$r=220M$ under the ingoing-wave boundary condition at the horizon. By fitting the far-region solution~\eqref{Farregionsol} with the numerical solution in the region of $200M\le r\le 220M$, we read off the coefficients~$A_+$ and $A_-$, and determine the response function~\eqref{responseF}. We have obtained the qualitatively same results for various upper limits of the integration and the locations of the fitting region with various widths. 

\begin{figure}[htbp]
\centering
\subfigure[Response function and the Love number]{
\includegraphics[scale=0.50]{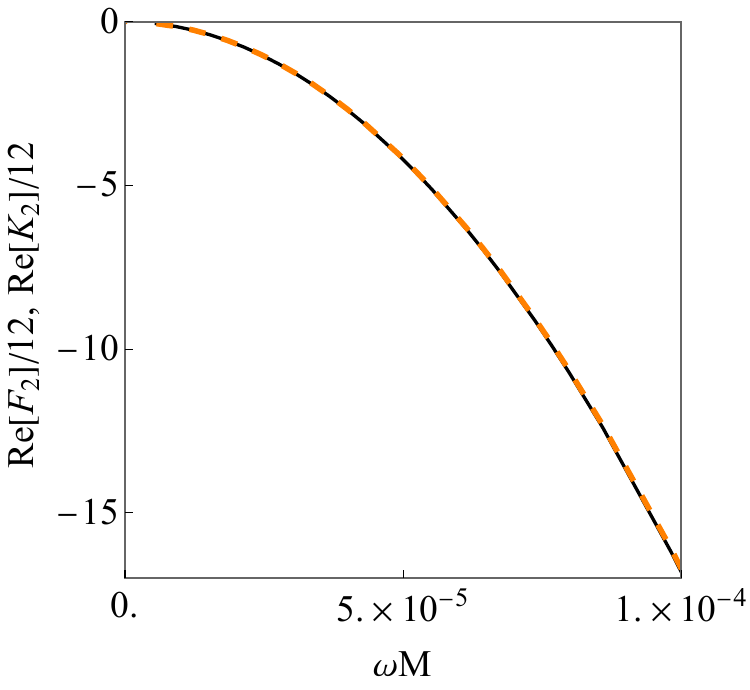}
\label{ReResponseFunction_Schwarzschild}}
\hspace{0.3cm}
\subfigure[Response function and the dissipation number]{
\includegraphics[scale=0.52]{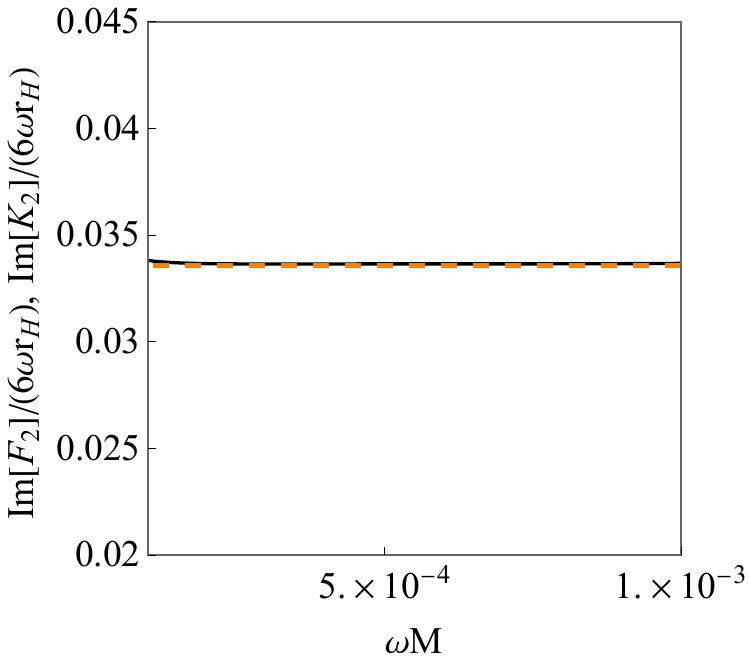}
\label{ImResponseFunction_Schwarzschild}}
\caption{({\it Left}) The functions~${\rm Re}[{\cal F}_2]/12$~(black solid line) and ${\rm Re}[{\cal K}_2]/12$~(orange dashed line), which correspond to the quadrupolar magnetic-type tidal Love number~$\kappa_2^B$ in the static limit~$\omega\to0$. ({\it Right}) the functions~${\rm Im}[{\cal F}_2]/(6\omega r_H)$~(black solid line) and ${\rm Im}[{\cal K}_2]/(6\omega r_H)$~(orange dashed line), which correspond to the quadrupolar magnetic-type dissipation number~$\nu_2^B$ in the limit~$\omega\to0$. The value of ${\rm Im}[{\cal F}_2]/(6\omega r_H)$ is almost constant~$0.0333$ in a good agreement with the analytic result from Eq.~\eqref{ImKell}}
\end{figure}

Figures~\ref{ReResponseFunction_Schwarzschild} and~\ref{ImResponseFunction_Schwarzschild} demonstrate the behavior of the functions~${\rm Re}[{\cal F}_2]/12$ and~${\rm Im}[{\cal F}_2]/(6\omega r_H)$, which correspond, respectively, to the quadrupolar magnetic-type tidal Love number~$\kappa_2^B$ and dissipation number~$\nu_2^B$ in the limit~$\omega\to0$. The overlapping of ${\rm Re}[{\cal F}_2]/12$~(black solid line) and ${\rm Re}[{\cal K}_2]/12$~(orange dashed line) means that the matching condition~\eqref{MatchingCondition} is indeed satisfied. In Fig.~\ref{ReResponseFunction_Schwarzschild}, we can see that the function~${\rm Re}[{\cal F}_2]/12$ approaches zero within the numerical error as $\omega M$ becomes sufficiently small, showing the vanishing of the Love number. This is consistent with the analytic consideration in the previous section. 
In Fig.~\ref{ImResponseFunction_Schwarzschild}, ${\rm Im}[{\cal F}_2]/(6\omega r_H)$ is almost constant~$0.0333$ in good agreement with the analytic result from Eq.~\eqref{ImKell}.

\subsection{Advantage of computation of the tidal Love and dissipation numbers from the response function}
The computation of a tidal response in terms of scattering waves allows one to overcome the difficulties stated in Sec.~\ref{Section:IntroductionAndSummary}. The evaluation of the tidal Love and dissipation numbers does not fully rely on the property of the hypergeometric function. As already seen in Eq.~\eqref{asymptoticPhiN}, the conventional manner relies on the analytic continuation between the fundamental solutions around the event horizon and around large distances in the near region. 

If one considers a tiny modification to the Schwarzschild background in the strong-field regime as in the following section, it is in general difficult to derive the exact solution in the near region in the form of a well-known function. Therefore, it is hard to obtain the coefficient corresponding to ${\cal K}_\ell$ in Eq.~\eqref{responseK}. On the other hand, perturbation fields in the far region are less affected by such modification near the horizon; therefore, a far-region solution can be still described by Eq.~\eqref{Farregionsol}. Here, the analytic continuation of $\ell$ from an integer to a generic number makes the distinction between the growing term in $r$ and the decaying one. To end, computing a tidal response boils down to the problem to evaluate the static limit of the response function~${\cal F}_\ell$ in Eq.~\eqref{responseF} once we obtain the values of $A_+$ and $A_-$ in the far-region solution~\eqref{Farregionsol} numerically, and then obtain the Love and dissipation numbers through the formulas for $\kappa_\ell^B$ and $\nu_\ell^B$ shown in Eqs.~\eqref{responseFandLoveNumbers} and~\eqref{responseFandDissipationNumbers}.

The gauge ambiguity is bypassed by computing the tidal Love and dissipation numbers in terms of the response function. This is because the Love and dissipation numbers as the property of sufficiently low-frequency scattering waves are determined under the boundary conditions that capture the physical property of a compact object in the strong-field regime and that of scattering waves in the weak-field regime~\cite{Chia:2020yla,Creci:2021rkz,Ivanov:2022qqt,Ivanov:2022hlo}. 

\section{Stability of tidal Love and dissipation numbers}
\label{Section:tidalresponseofdeformedSchwarzschild}

In this section, we discuss the effect of a small deformation of the effective potential in the Regge-Wheeler equation on the tidal Love and dissipation numbers by using the formulas~\eqref{responseFandLoveNumbers} and~\eqref{responseFandDissipationNumbers}. We then show that a small deformation of the potential in the form of a Gaussian small bump  gives rise to non-zero values sensitive to the property of the bump and its deviation from zero is much larger than the scale of the bump, exhibiting destabilization of the Love numbers. The dissipation numbers, on the other hand, are stable. 
We also analyze stability against a potential modification that consists of combination of Gaussian bump and dip. In Appendix~\ref{Appendix:HCOcase}, we discuss stability of tidal Love and dissipation numbers in the case where one imposes a reflective boundary condition at a radius slightly outside the Schwarzschild radius as a proxy of horizonless compact objects.

\subsection{Modeling}
\begin{figure}[htbp]
\centering
\includegraphics[scale=0.68]{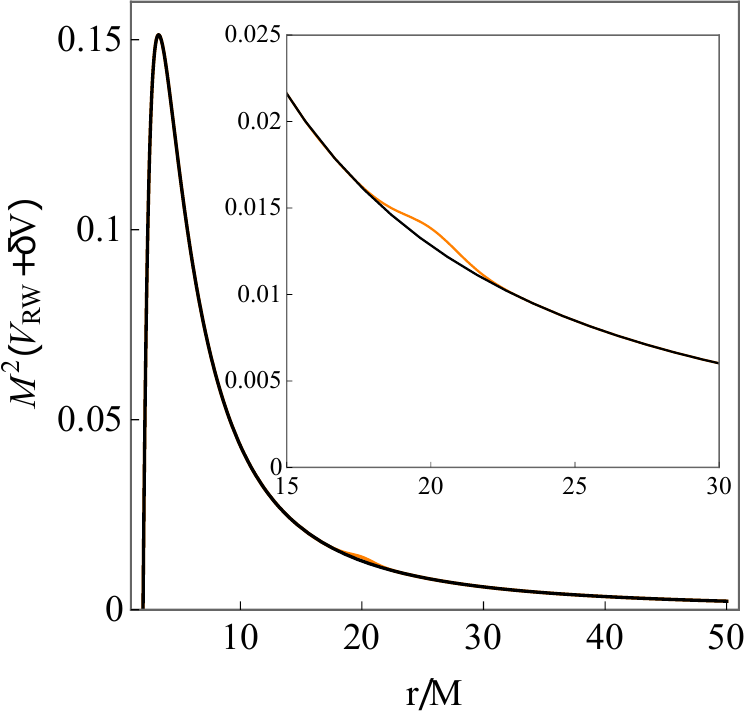}
\caption{The unperturbed potential~$\delta V=0$ (black line) and deformed potential by the Gaussian bump~\eqref{Gaussianbump} with $h=10^{-3}/M^2$, $\sigma=M$, and $a=20 M$~(orange line). Here, $V_{\rm RW}$ is the Regge-Wheeler potential; $V_{\rm RW}=(1-2M/r)[\ell(\ell+1)/r^2-6M/r^3]$. The inset shows the enlargement around the bump.}
\label{DeformedPotential}
\end{figure}
Let us consider the Regge-Wheeler equation with its effective potential slightly deformed:
\begin{equation}
\label{RWeqwithbump}
\left(1-\frac{2M}{r}\right)\frac{d}{dr}\left[\left(1-\frac{2M}{r}\right)\frac{d\Phi}{dr}\right]+\left\{\omega^2-\left(1-\frac{2M}{r}\right)\left[\frac{\ell(\ell+1)}{r^2}-\frac{6M}{r^3}\right]-\delta V\right\}\Phi=0.
\end{equation}
As the potential deformation, we here consider a Gaussian small bump,
\begin{equation}
\label{Gaussianbump}
\delta V=
h\exp\left[-\dfrac{\left(r-a\right)^2}{2\sigma^2}\right],
\end{equation}
which is characterized by its height~$h (\ll1/M^2)$, width~$\sigma$, and location of the peak~$a$. The explicit form of the deformed potential, as an example, is presented in Fig.~\ref{DeformedPotential} for the quadrupolar mode~$\ell=2$.

A small bump can be realized by some spherically symmetric local matter distribution~\cite{Cheung:2021bol}. Such a configuration may not be necessarily realistic in astrophysical situations but can be expected to provide a certain insight into the environmental effect on the tidal Love numbers by matter surroundings including accretion disks~\cite{Cardoso:2019upw}.

\subsection{Effect of a Gaussian bump on the tidal Love and dissipation numbers}
\begin{figure}[htbp]
\centering
\subfigure[Response function and the Love number]{
\includegraphics[scale=0.52]{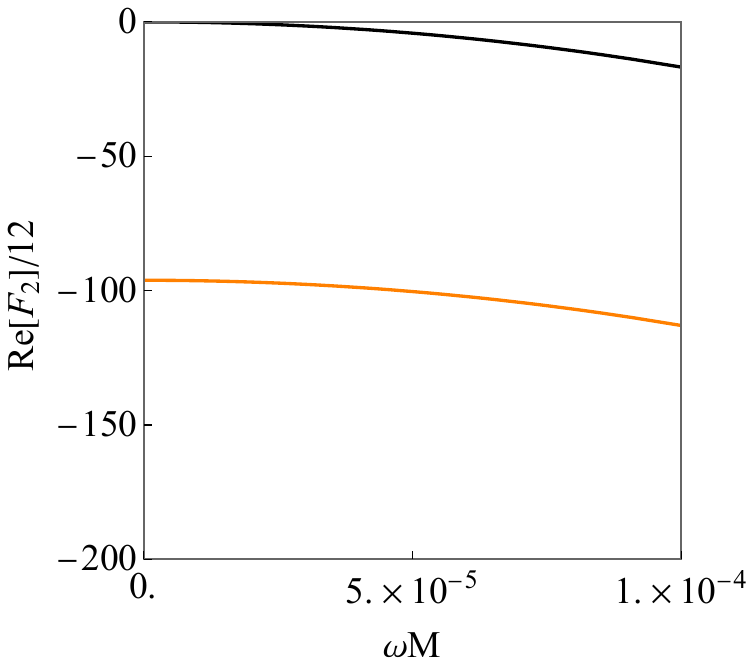}
\label{ReResponseFunction_Gaussian1}}
\hspace{0.3cm}
\subfigure[Response function and the dissipation number]{
\includegraphics[scale=0.53]{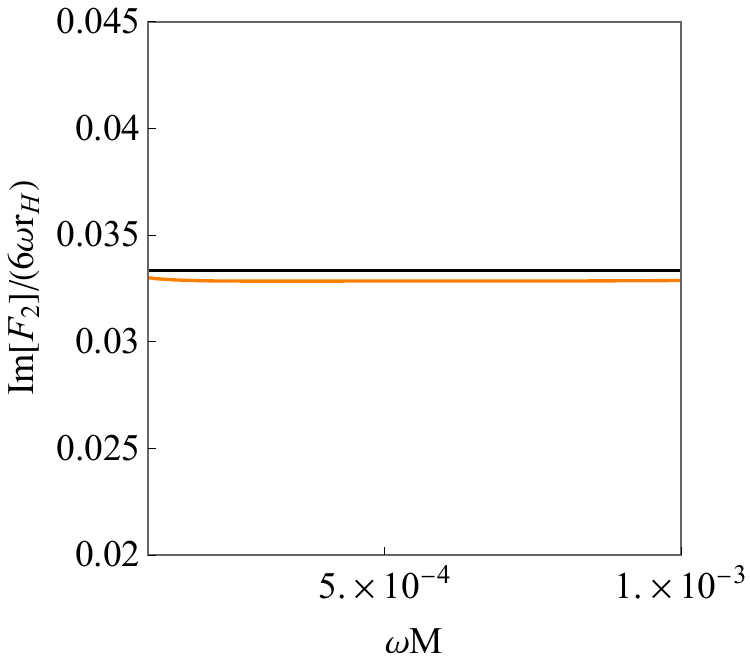}
\label{ImResponseFunction_Gaussian1}}
\caption{The functions~${\rm Re}[{\cal F}_2]/12$~(the left panel) and~${\rm Im}[{\cal F}_2]/(6\omega r_H)$~(the right panel) which correspond to, respectively, the quadrupolar magnetic-type tidal Love number~$\kappa_2^B$ and dissipation number~$\nu_2^B$ in the limit~$\omega \to0$, showing $\kappa_2^B\simeq -96$ and $\nu_2^B\simeq 0.0328$, in the presence of the Gaussian bump with $h=10^{-3}/M^2$, $\sigma=M$, and $a=20M$~(orange line). The black line corresponds to the unperturbed potential case~($\delta V=0$) and is the same as those in Figs.~\ref{ReResponseFunction_Schwarzschild} and~\ref{ImResponseFunction_Schwarzschild}.}
\end{figure}

\begin{figure}[htbp]
\centering
\subfigure[For various locations]{
\includegraphics[scale=0.51]{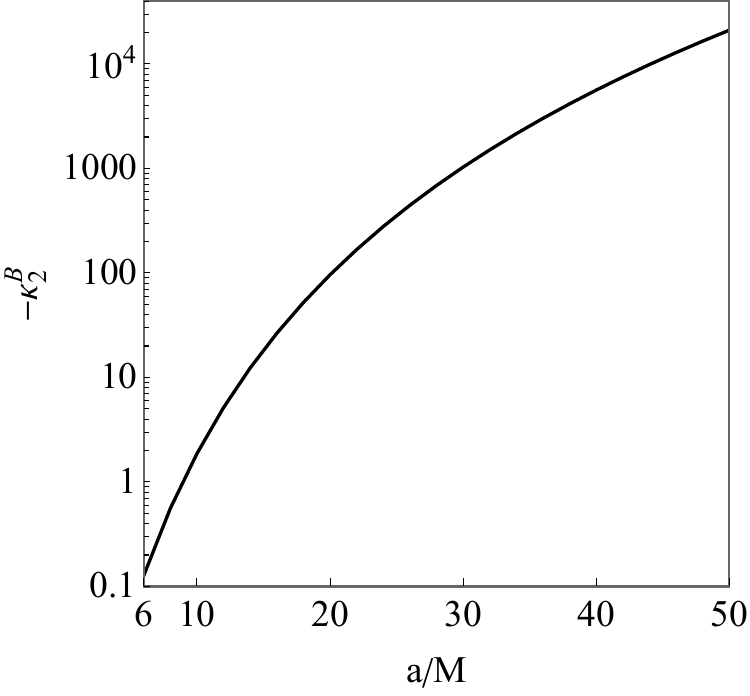}
\label{LimitOfReResponseFunction_Gaussian1_a}}
\hspace{0.3cm}
\subfigure[For various heights]{
\includegraphics[scale=0.54]{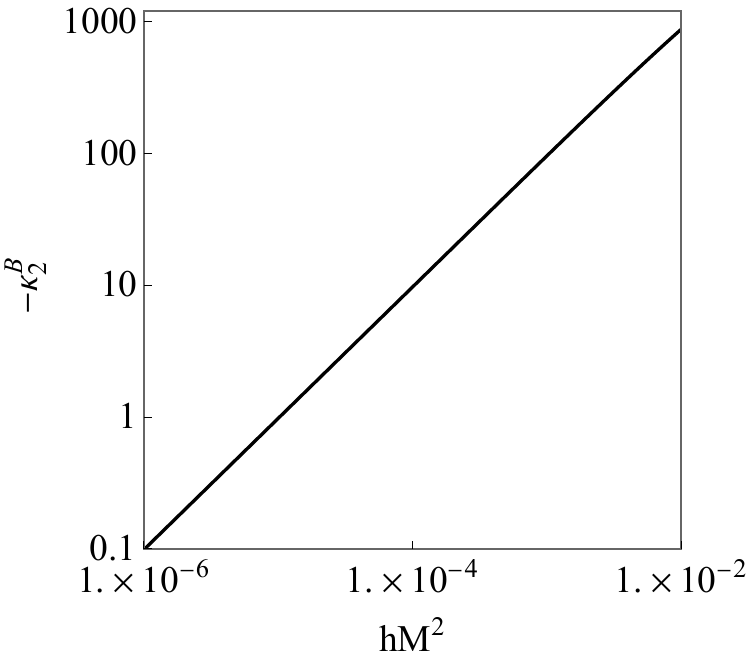}
\label{LimitOfReResponseFunction_Gaussian1_h}}
\caption{The quadrupolar magnetic-type tidal Love numbers with the opposite sign, i.e., $-\kappa_2^B$, for various locations~$a$ and heights~$h$ of the Gaussian bump~\eqref{Gaussianbump}, ({\it Left}) for the locations from the radius of the innermost stable circular orbit~$a=6M$ to $a=50M$ with $h=10^{-3}/M^2$ and $\sigma=M$, and ({\it Right}) for the heights from $h=10^{-6}/M^2$ to $h=10^{-2}/M^2$ with $a=20M$ and $\sigma=M$.}
\end{figure}

We first discuss the behavior of the response function~${\cal F}_\ell$ in Eq.~\eqref{responseF}.
Figures~\ref{ReResponseFunction_Gaussian1} and~\ref{ImResponseFunction_Gaussian1} give the functions~${{\rm Re}[\cal F}_2]/12$ and~${\rm Im}[{\cal F}_2]/(6\omega r_H)$. The values on the vertical axis correspond to the quadrupolar magnetic-type tidal Love number~$\kappa_2^B$ and dissipation number~$\nu_2^B$, respectively. Figure~\ref{ReResponseFunction_Gaussian1} shows that the tidal Love number takes a negative value due to the small modification of the effective potential as $\kappa_2^B\simeq -96$. This implies that a tiny deviation from the exact vacuum environment gives a nonzero Love number whose deviation from zero is much larger than the scale of the small potential modification. On the other hand, Fig.~\ref{ImResponseFunction_Gaussian1} shows $\nu_2^B\simeq 0.0328$ even in the presence of the bump, implying that the modification has less impact on the imaginary part of the response function.

We next compute the Love and dissipation numbers from the formulas~\eqref{responseFandLoveNumbers} and~\eqref{responseFandDissipationNumbers} in the current deformed system. Figures~\ref{LimitOfReResponseFunction_Gaussian1_a} and~\ref{LimitOfReResponseFunction_Gaussian1_h} give the quadrupolar magnetic-type tidal Love numbers with the opposite sign, i.e.,~$-\kappa_2^B$, for various values of the location~$a$ and height~$h$, respectively, of the Gaussian bump~\eqref{Gaussianbump}. It is demonstrated that the tidal Love number blows up nonlinearly/linearly with increasing location/height. This implies that the Love numbers are sensitive to the properties of small modifications to the effective potential, exhibiting destabilization of the tidal Love numbers. This destabilization comes from the property of the Love number of the Gaussian bump as seen in Appendix~\ref{Appendix:BHplusBumpinMinkowski}. The nonlinear blow-up with respect to the location is consistent with the analytic result for a thin shell in Ref.~\cite{Cardoso:2019upw}. We comment that the absolute value of the Love number, i.e., $|\kappa_2^B|$, grows as $\sigma$ increases, indicating that a wider bump renders a larger tidal deformation. 

\begin{figure}[htbp]
\centering
\subfigure[For various locations]{
\includegraphics[scale=0.5]{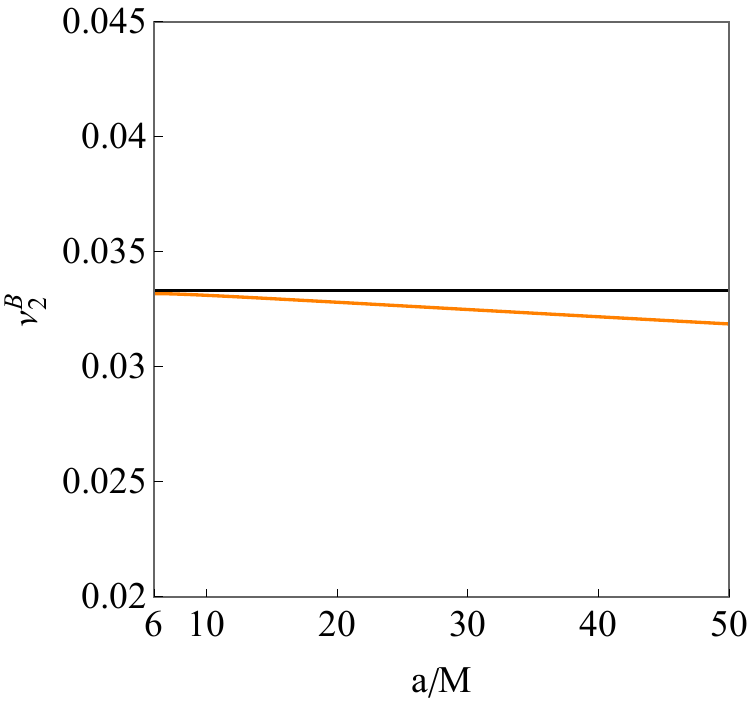}
\label{LimitOfImResponseFunction_Gaussian1_a}}
\hspace{0.3cm}
\subfigure[For various heights]{
\includegraphics[scale=0.54]{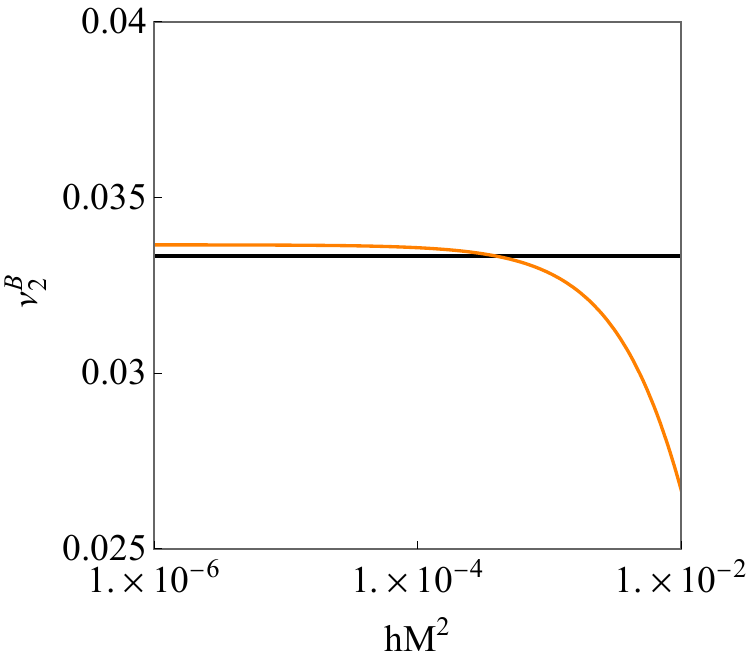}
\label{LimitOfImResponseFunction_Gaussian1_h}}
\caption{The quadrupolar magnetic-type dissipation numbers~$\nu_2^B$ for various locations~$a$ and heights~$h$ of the Gaussian bump~\eqref{Gaussianbump}~(orange line) and those of the Schwarzschild black hole without the bump, i.e., $\nu_2^B|_{\delta V=0}$ ($= 0.0333$),~(black line). The parameter set is the same as that in Figs.~\ref{LimitOfReResponseFunction_Gaussian1_a} and~\ref{LimitOfReResponseFunction_Gaussian1_h}. }
\end{figure}
Figures~\ref{LimitOfImResponseFunction_Gaussian1_a} and~\ref{LimitOfImResponseFunction_Gaussian1_h} show that the quadrupolar magnetic-type dissipation number~$\nu_2^B$ is smaller for a bump at a more distant location and with a higher height. This implies that a more distant or higher bump obstructs more the absorption effect of the black hole. The relative differences from $\nu_2^B|_{\delta V=0}$ ($=0.0333$) are kept within $20\%$ in the current parameter domain. The less sensitivity to the property of the modification means that the dissipation numbers are stable.

Why does such destabilization occur for tidal Love numbers and does not for dissipation numbers? The answer is as follows: a tidal response of a modified system consists of that of the black hole and of the outer Gaussian bump. In other words, the tidal Love and dissipation numbers in a composite system are approximately determined by the linear combination of those of each component of the system. 
A Gaussian bump in the Minkowski spacetime has non-zero Love numbers and zero dissipation numbers (see Appendix~\ref{Appendix:BHplusBumpinMinkowski}). In fact, the former takes a value close to that of the Schwarzschild black hole with the same Gaussian bump and shares the almost same dependence on the parameters of the bump~(see Figs.~\ref{LimitOfReResponseFunction_GaussianinMinkowski_a} and~\ref{LimitOfReResponseFunction_GaussianinMinkowski_h}). The latter has less impact on the dissipation numbers of the black hole. The above result on the tidal Love number supports the claim of Ref.~\cite{Cardoso:2019upw}.

We have checked that a negative~$h$, i.e., a ``dip'', leads to a positive tidal Love number and a positive dissipation number, which is larger than $\nu_2^B|_{\delta V=0}$ ($=0.0333$). They are more for a deeper Gaussian dip. It is worth noting that the sign of the tidal Love number is governed solely by that of $h$, which controls whether the Gaussian modification is repulsive or attractive for scattering waves at sufficiently low frequencies. 

The behavior of the dissipation number can be interpreted in terms of the scattering. The nonzero dissipation number arises from the dissipation of sufficiently low-frequency scattering waves. In the presence of a Gaussian bump, the obstruction of absorption into the event horizon results in the decrease of the dissipation number. A Gaussian dip, on the other hand, leads to more attenuation of outgoing spherical waves, resulting in larger dissipation numbers. Note that the dissipation number never changes its sign while varying the height of the Gaussian bump because the modification plays a role only in obstructing the absorption of waves into the black hole horizon. The negative dissipation number means the existence of outgoing waves from the horizon even though we imposed the ingoing-wave boundary condition, contradicting physical intuition.

\subsection{Stability against combination of the potential deformation}
\label{Section:CaseOftwoGaussian}
\begin{figure}[htbp]
\centering
\includegraphics[scale=0.68]{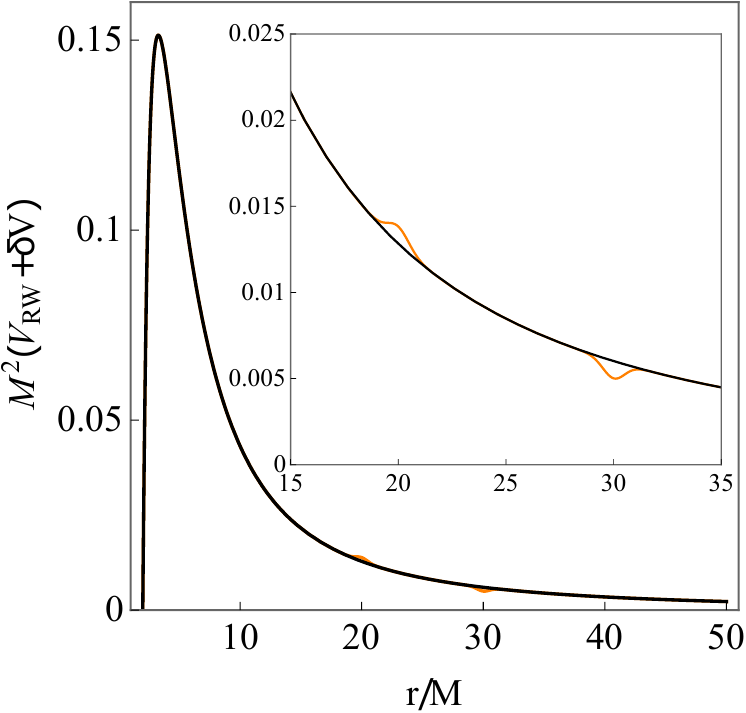}
\caption{The unperturbed potential~$\delta V_1=\delta V_2=0$ (black line) and deformed potential by the Gaussian bump and dip in Eq.~\eqref{Gaussianbumpdip} with $h_1=-h_2=10^{-3}/M^2$, $\sigma_1=\sigma_2=M$, $a_1=20 M$, and $a_2=a_1+10M$~(orange line). Here, $V_{\rm RW}$ is the Regge-Wheeler potential, $V_{\rm RW}=(1-2M/r)[\ell(\ell+1)/r^2-6M/r^3]$. The inset shows the enlargement around the bump and dip.}
\label{DeformedPotential2}
\end{figure}
We consider a system where a potential deformation comes from combination of any two of Gaussian bump(s) and/or dip(s),
\begin{align}
& \left(1-\frac{2M}{r}\right)\frac{d}{dr}\left[\left(1-\frac{2M}{r}\right)\frac{d\Phi}{dr}\right]
\cr & 
+\left\{\omega^2-\left(1-\frac{2M}{r}\right)\left[\frac{\ell(\ell+1)}{r^2}-\frac{6M}{r^3}\right]-\delta V_1-\delta V_2\right\}\Phi=0,
\end{align}
where
\begin{equation}
\begin{split}
\label{Gaussianbumpdip}
\delta V_1&=h_1\exp\left[-\dfrac{\left(r-a_1\right)^2}{2\sigma_1^2}\right],\\
\delta V_2&=h_2\exp\left[-\dfrac{\left(r-a_2\right)^2}{2\sigma_2^2}\right].
\end{split}
\end{equation}
Here, we have assumed $|h_1|\ll1/M^2$, $|h_2|\ll1/M^2$, and that the second modification~$\delta V_2$ is located at a larger radius than the first one~$\delta V_1$, i.e., $a_2\ge a_1$. The explicit form of the deformed potential is depicted in Fig.~\ref{DeformedPotential2} for the quadrupolar mode~$\ell=2$.
An example of such a bumpy deformation around the Regge-Wheeler potential, which is caused by a local matter distribution, is presented in Appendix~\ref{Appendix:potentialbymatterfield}~(see Fig.~\ref{DeformedPotential_3}). 
\begin{figure}[htbp]
\centering
\subfigure[For various locations]{
\includegraphics[scale=0.5]{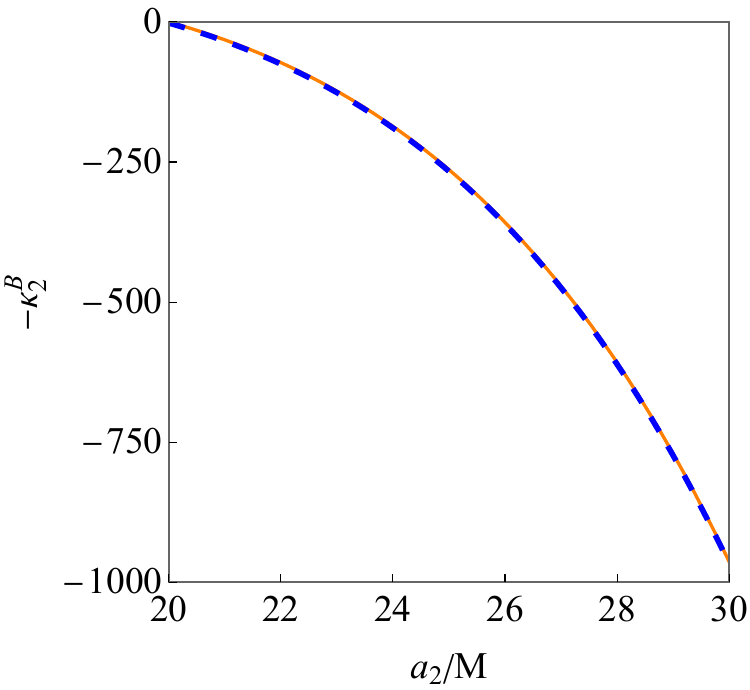}
\label{LimitOfReResponseFunction_Gaussian2_a}}
\hspace{0.3cm}
\subfigure[For various depths]{
\includegraphics[scale=0.54]{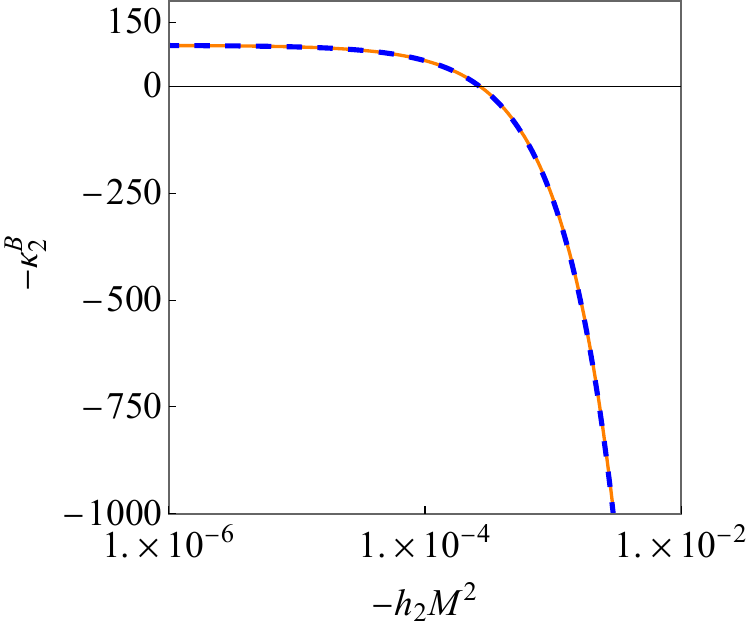}
\label{LimitOfReResponseFunction_Gaussian2_h}}
\caption{The quadrupolar magnetic-type tidal Love numbers with the opposite sign,~i.e.,~$-\kappa_2^B$, for various locations~$a_2$ and depths~$h_2$ of the Gaussian dip~$\delta V_2$ in Eq.~\eqref{Gaussianbumpdip} with a fixed Gaussian bump~$\delta V_1$~(orange solid line) and the values of $-(\kappa_2^B|_{\delta V_2=0}+\kappa_2^B|_{\delta V_1=0})$~(blue dashed line), ({\it Left}) for the locations from $a_2=a_1(=20M)$ to $a_2=a_1+10M$ with $h_2=-h_1=-10^{-3}/M^2$ and $\sigma_2=\sigma_1=M$, and ({\it Right}) for the depth from $-h_2=10^{-6}/M^2$ to $-h_2=10^{-2}/M^2$ with $a_2=a_1+5M$ and $\sigma_2=\sigma_1=M$.
The parameter set of $\delta V_1$ is the same as those in Fig.~\ref{ReResponseFunction_Gaussian1}.
}
\end{figure}

\begin{figure}[htbp]
\centering
\subfigure[For various locations]{
\includegraphics[scale=0.5]{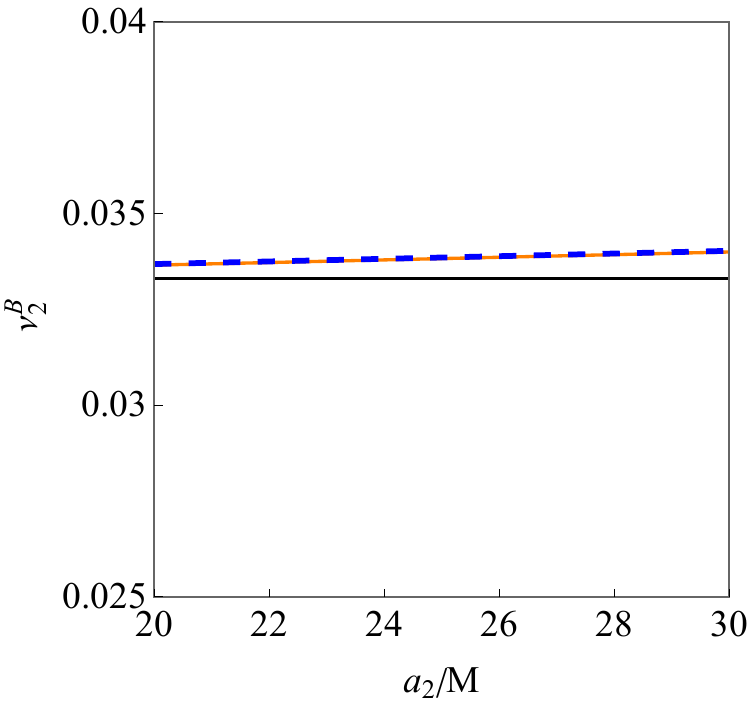}
\label{LimitOfImResponseFunction_Gaussian2_a}}
\hspace{0.3cm}
\subfigure[For various depths]{
\includegraphics[scale=0.53]{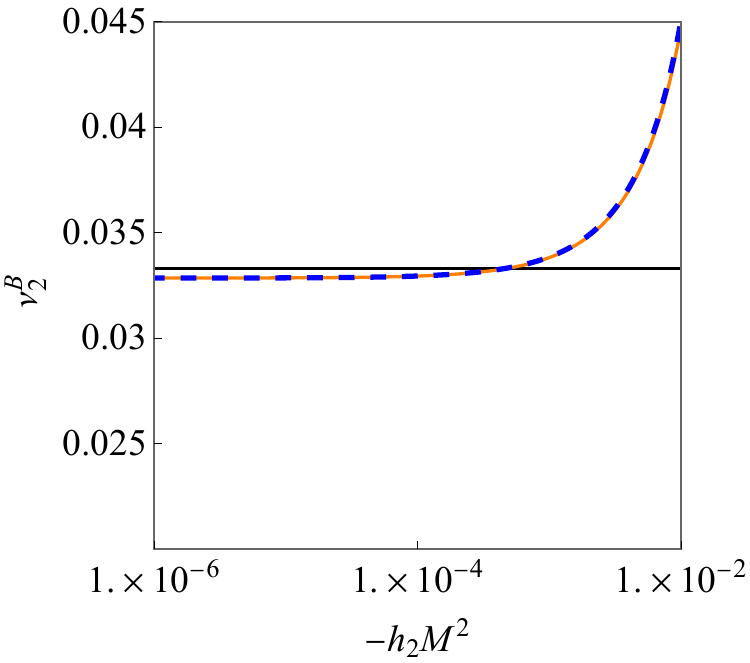}
\label{LimitOfImResponseFunction_Gaussian2_h}}
\caption{The quadrupolar magnetic-type dissipation numbers~$\nu_2^B$ for various locations~$a_2$ and depths~$h_2$ of the Gaussian dip~$\delta V_2$ in Eq.~\eqref{Gaussianbumpdip} with a fixed Gaussian bump~$\delta V_1$~(orange solid line) and the values of $\nu_2^B|_{\delta V_2=0}+\nu_2^B|_{\delta V_1=0}-\nu_2^B|_{\delta V_1=\delta V_2=0}$~(blue dashed line). The last subtraction eliminates the overlapping of the contribution from the purely black hole case, $\nu_2^B|_{\delta V_1=\delta V_2=0}=0.0333$, in the combination of the first two components. The horizontal black solid line corresponds to $\nu_2^B|_{\delta V_1=\delta V_2=0}=0.0333$. The parameter set of $\delta V_1$ and $\delta V_2$ is, respectively, the same as those in Fig.~\ref{ReResponseFunction_Gaussian1} and those in Figs.~\ref{LimitOfReResponseFunction_Gaussian2_a} and~\ref{LimitOfReResponseFunction_Gaussian2_h}. }
\end{figure}

To be specific, we consider combination of a Gaussian bump~$\delta V_1$ with $h_1>0$ and dip~$\delta V_2$ with $h_2<0$ in the following. For other cases where the inner and outer modifications are the Gaussian dip and bump respectively, or both are bumps or dips, the following results remain qualitatively the same. 

Figures~\ref{LimitOfReResponseFunction_Gaussian2_a} and~\ref{LimitOfReResponseFunction_Gaussian2_h}, respectively, show the quadrupolar magnetic-type tidal Love number with the opposite sign, i.e., $-\kappa_2^B$, as a function of the location and depth of the Gaussian dip~$\delta V_2$ with the Gaussian bump~$\delta V_1$ being fixed with the same parameter set as in Figs.~\ref{ReResponseFunction_Gaussian1} and~\ref{ImResponseFunction_Gaussian1}. Notice that the value in the presence of both $\delta V_1$ and $\delta V_2$~(orange solid line) is almost the same as the linear combination of those in case where each of the Gaussian bump~$\delta V_1$ or the Gaussian dip~$\delta V_2$ is present, i.e., $-(\kappa_2^B|_{\delta V_2=0}+\kappa_2^B|_{\delta V_1=0})$~(blue dashed line). This implies that the tidal Love number in a composite system is mostly determined by the linear combination of the Love number of each component in the system. It is worth mentioning that in Fig.~\ref{LimitOfReResponseFunction_Gaussian2_h}, the tidal Love number can vanish for a specific modification.

Figures~\ref{LimitOfImResponseFunction_Gaussian2_a} and~\ref{LimitOfImResponseFunction_Gaussian2_h} present the quadrupolar magnetic-type dissipation number~$\nu_2^B$ as a function of the location~$a_2$ and depth~$h_2$ of the Gaussian dip~$\delta V_2$, respectively. The value of $\nu_2^B$~(orange solid line) is almost the same as $\nu_2^B|_{\delta V_2=0}+\nu_2^B|_{\delta V_1=0}-\nu_2^B|_{\delta V_1=\delta V_2=0}$~(blue dashed line), where the last subtraction eliminates the overlap of the contribution from the purely black hole case, $\nu_2^B|_{\delta V_1=\delta V_2=0}=0.0333$, in the combination of the first two components. This implies that the dissipation number is approximately determined by the linear combination of the number of each component in the system. Since its deviation from $\nu_2^B|_{\delta V_1=\delta V_2=0}$ is small, we conclude that the dissipation numbers are still stable even in the presence of another modification.

\section{Discussion}
\label{Section:discussion}

We discuss the astrophysical implication and theoretical application of the results shown in the previous sections.

\subsection{Astrophysical implications}
In the previous sections, it is shown that a Schwarzschild black hole acquires non-zero tidal Love numbers due to the presence of a Gaussian small modification
and their values are sensitive to the property of the deformation. This means that, even if non-zero Love number is measured in future gravitational-wave observations, we cannot immediately conclude deviation of the underlying theory of gravity from General Relativity without careful consideration on environmental effects. From another viewpoint, a non-zero Love number allows us to catch a glimpse of the extreme property of matter
fields around a black hole through gravitational-wave observations. In yet another context, the destabilization of the Love numbers may hinder constraining the matter equation of state in neutron stars because the destabilization occurs even for horizonless compact objects including a neutron star as seen in Appendix~\ref{Appendix:HCOcase}. 

The dissipation numbers are stable for small modifications. Therefore, in the context of a test of quantum corrections in the strong-field regime~\cite{Addazi:2018uhd,Cardoso:2019rvt}, quantifying the existence of the event horizon is not spoiled even with deformation of the potential due to the presence of a matter field.

In an inspiraling binary, the environmental effect causing a potential deformation varies with time as the orbital separation decreases. Therefore, the tidal response induced by the environment is not constant~(see e.g., Ref.~\cite{DeLuca:2022xlz}). On the other hand, the tidal Love numbers arising from the modification in theories of gravity remains constant. The tidal response measured with gravitational-wave observations will be approximately determined by the linear combination of the time-varying part and the constant part. Thus, the extraction of the constant component from the time-varying tidal response will be an important step in testing theories of gravity in the strong-field regime.

\subsection{Theoretical application: power-law correction to the effective potential}
Toward testing theories of gravity in the strong-gravity regime within linear perturbation theory, ``parametrized'' formalism is expected to be useful~\cite{Cardoso:2019mqo,McManus:2019ulj,Volkel:2022aca}:
\begin{equation}
\label{parameterziedV}
\left(1-\frac{r_H}{r}\right)\frac{d}{dr}\left[\left(1-\frac{r_H}{r}\right)\frac{d\Phi}{dr}\right]+\left\{\omega^2-\left(1-\frac{r_H}{r}\right)\left[\frac{\ell(\ell+1)}{r^2}-\frac{3r_H}{r^3}\right]-\delta V\right\}\Phi=0,
\end{equation}
with 
\begin{equation}
\delta V=\frac{1}{r_H^2}\left(1-\frac{r_H}{r}\right)\sum_{j=0}^{\infty}\alpha_j \left(\frac{r_H}{r}\right)^j,
\end{equation}
where $|\alpha_j|\ll (1+1/j)^{j}(j+1)$~\cite{Cardoso:2019mqo}. Choosing the coefficient~$\alpha_j$ appropriately, one can reconstruct the effective potential for linear odd-parity gravitational perturbations around a static and spherically symmetric black hole in a specific theory~(see Ref.~\cite{Cardoso:2019mqo}). As another application, the parametrized formalism may also be of use for modeling the deviation from a Schwarzschild background due to a continuous matter distribution~\cite{Barausse:2014tra}.

The analysis in terms of the Gaussian bump or dip in Sec.~\ref{Section:tidalresponseofdeformedSchwarzschild} gives various suggestions for the properties of tidal Love and dissipation numbers in a deformed system~\eqref{parameterziedV}. First, the dissipation numbers take values close to those in the purely Schwarzschild case at least for small~$|\alpha_j|$. Second, the Love and dissipation numbers for multiple corrections can be interpreted as the linear combination of those of single power-law corrections and those of the Schwarzschild black hole approximately. Third, the Love number can vanish even with the corrections. Fourth, different parameter sets of $\alpha_j$ and $j$ can give an identical Love number. For the Gaussian bump, this degeneracy can be seen in Figs.~\ref{LimitOfReResponseFunction_Gaussian1_a} and~\ref{LimitOfReResponseFunction_Gaussian1_h}. Finally, for multiple corrections, a lower-order contribution of $j$ dominates over a higher-order one in determining the Love numbers if the absolute value of the coefficient of the lower-order one is larger than or comparable with those of the higher-order one.

\acknowledgments
The authors wish to express their cordial gratitude to Prof. Takahiro Tanaka, the Leader of Innovative Area Grants-in-Aid for Scientific Research ``Gravitational wave physics and astronomy: Genesis'', for his continuous interest and encouragement. The authors would also like to thank Vitor Cardoso, Kazumi Kashiyama, and Shijun Yoshida for useful comments. The authors acknowledge Chams Gharib Ali Barura, Shinji Mukohyama, Kazufumi Takahashi, and Vicharit Yingcharoenrat for a lot of critical comments and fruitful discussions on the analysis in Sec.~\ref{Section:tidalresponseofSchwarzschildBH}. This research is supported by Grants-in-Aid for Scientific Research (TK and KO: 17H06360, HN: 17H06358, 21H01082, 21K03582, 23K03432, KO: 17H01102, 17H02869, 22H00149) from the Japan Society for the Promotion of Science. TK thanks for support by VILLUM FONDEN (grant no. 37766), by the Danish Research Foundation, and under the European Union’s H2020 ERC Advanced Grant ``Black holes: gravitational engines of discovery'' grant agreement no. Gravitas–101052587. KO acknowledges support from the Amaldi Research Center funded by the MIUR program ``Dipartimento di Eccellenza'' (CUP:B81I18001170001).

\appendix
\section{Tidal response of a Gaussian bump in the Minkowski spacetime}
\label{Appendix:BHplusBumpinMinkowski}
\begin{figure}[htbp]
\centering
\subfigure[For various locations]{
\includegraphics[scale=0.5]{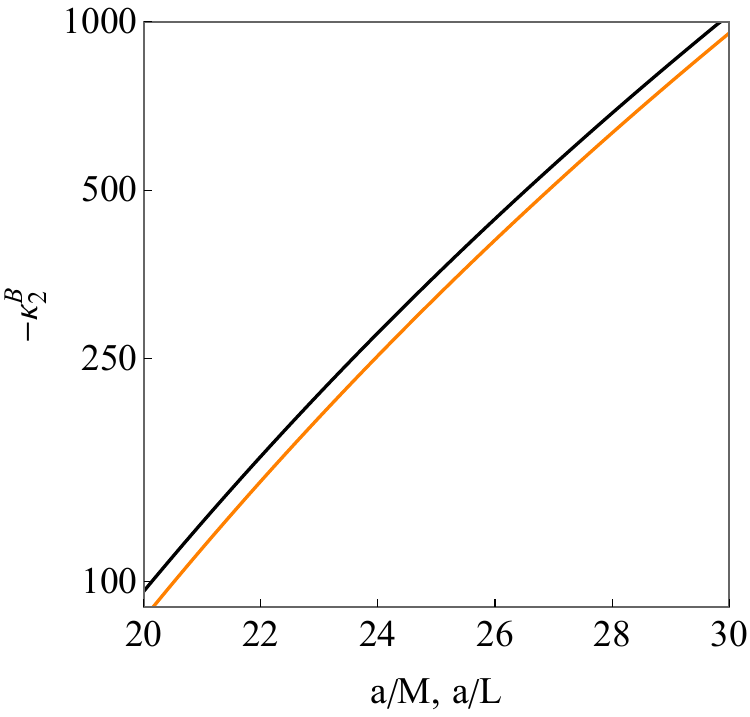}
\label{LimitOfReResponseFunction_GaussianinMinkowski_a}}
\hspace{0.3cm}
\subfigure[For various heights]{
\includegraphics[scale=0.53]{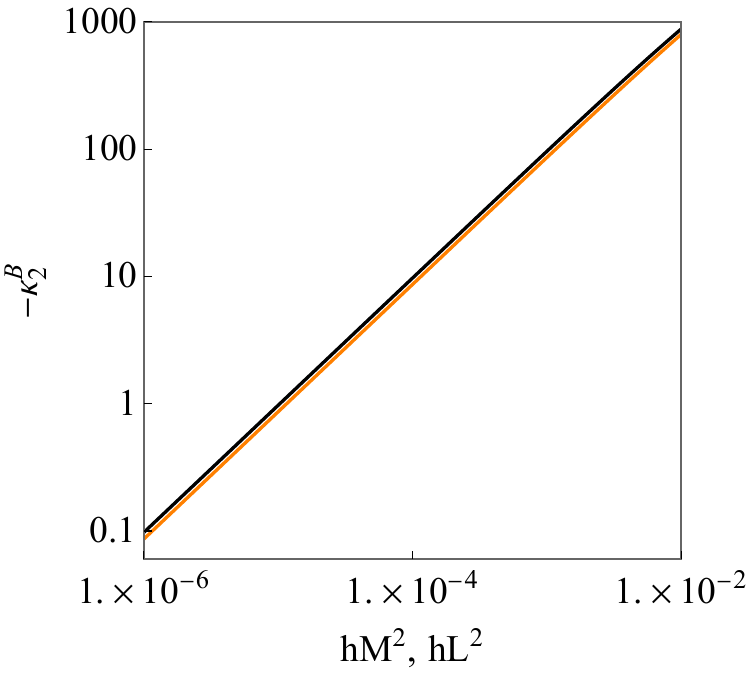}
\label{LimitOfReResponseFunction_GaussianinMinkowski_h}}
\caption{The quadrupolar magnetic-type tidal Love numbers with the opposite sign, i.e., $-\kappa_2^B$, for various locations~$a$ and heights~$h$ of the Gaussian bump~\eqref{Gaussianbump}. The black and orange  lines correspond to the values of the Schwarzschild black hole with the Gaussian bump and those of the Gaussian bump in the Minkowski spacetime, respectively. Here, $L$ is a length scale of the Gaussian bump in the Minkowski spacetime. ({\it Left}) for the locations from $a=20M=20L$ to $a=30M=30L$ with $h=10^{-3}/M^2=10^{-3}/L^2$ and $\sigma=M=L$. ({\it Right}) for the heights from $h=10^{-6}/M^2=10^{-6}/L^2$ to $h=10^{-2}/M^2=10^{-2}/L^2$ with $a=20M=20L$ and $\sigma=M=L$. }
\end{figure}

We discuss a tidal response of a Gaussian bump in the Minkowski spacetime. The bump~\eqref{Gaussianbump} is introduced in Eq.~\eqref{RWeqwithbump} with $M=0$:
\begin{equation}
\label{PoissonEqforbump}
\frac{d^2\Phi}{dr^2}+\left[\omega^2-\frac{\ell(\ell+1)}{r^2}+\delta V\right]\Phi=0.
\end{equation}
If $\delta V=0$, we have an analytic solution regular at the origin~$r=0$, i.e.,
\begin{equation}
\label{regularsol}
\Phi=r^{1/2}J_{\ell+1/2}\left( \omega r \right),
\end{equation}
where $J_{\ell+1/2}$ is the Bessel function of the first kind. We obtain the response function from a numerical solution which is obtained by integrating Eq.~\eqref{PoissonEqforbump} from the origin to large distances under the boundary condition~\eqref{regularsol} at $r=0$.

Figures~\ref{LimitOfReResponseFunction_GaussianinMinkowski_a} and~\ref{LimitOfReResponseFunction_GaussianinMinkowski_h} give the quadrupolar magnetic-type tidal Love numbers for various location with the opposite sign, i.e., $-\kappa_2^B$, and heights of the Gaussian bump, demonstrating that those of the bump in the Minkowski spacetime and those in the Schwarzschild spacetime have close values and share qualitatively the almost same behavior for the property of the potential deformation. We have checked that the relative difference of them is at most $10\%$. This implies that the tidal Love numbers of a Schwarzschild black hole with a Gaussian bump are mostly determined by the property of the bump.

It is found that the dissipation number~$\nu_2^B$ of the Gaussian bump in the Minkowski spacetime is a quite smaller value than unity, meaning its vanishing. This implies that the Gaussian bump has less impact on the dissipation numbers of a Schwarzschild black hole.

\section{Potential deformation by a local matter distribution}
\label{Appendix:potentialbymatterfield}

We here construct a static and spherically symmetric black hole solution with an anisotropic matter field, and then derive an effective potential for an odd-parity linear gravitational perturbation. In spherical polar coordinates~$(t,r,\theta,\varphi)$, a line element of a static and spherically symmetric spacetime is given by
\begin{equation}
\label{sphericallysymspacetime}
ds^2=-A(r)dt^2+\left(1-\frac{2m(r)}{r}\right)^{-1}dr^2+r^2\left(d\theta^2+\sin^2\theta d\varphi^2\right),
\end{equation}
where $A$ and $m$ are functions of the areal radius~$r$. The function~$m$ is the Misner-Sharp mass~\cite{PhysRev.136.B571,Hayward:1994bu}. We introduce a stationary system consisting of many gravitating masses which are assumed to be anisotropic and have only tangential pressure in the angular directions without the radial pressure. The energy-momentum tensor is given by
\begin{equation}
T_{\mu}{}^\nu={\rm diag}[-\rho(r),0,P_t(r),P_t(r)].
\end{equation}
The Bianchi identity gives the relation between the local energy density~$\rho$ and the pressure~$P_t$:
\begin{equation}
\label{Prhorelation}
P_t=\frac{\rho r}{4A}\frac{dA}{dr}.
\end{equation}
In addition, the~$(r,r)$-component of the Einstein equation leads to a constraint:
\begin{equation}
\label{constraint}
\frac{r}{2}\frac{dA}{dr}=\frac{m}{r-2m}.
\end{equation}
\begin{figure}[htbp]
\centering
\includegraphics[scale=0.68]{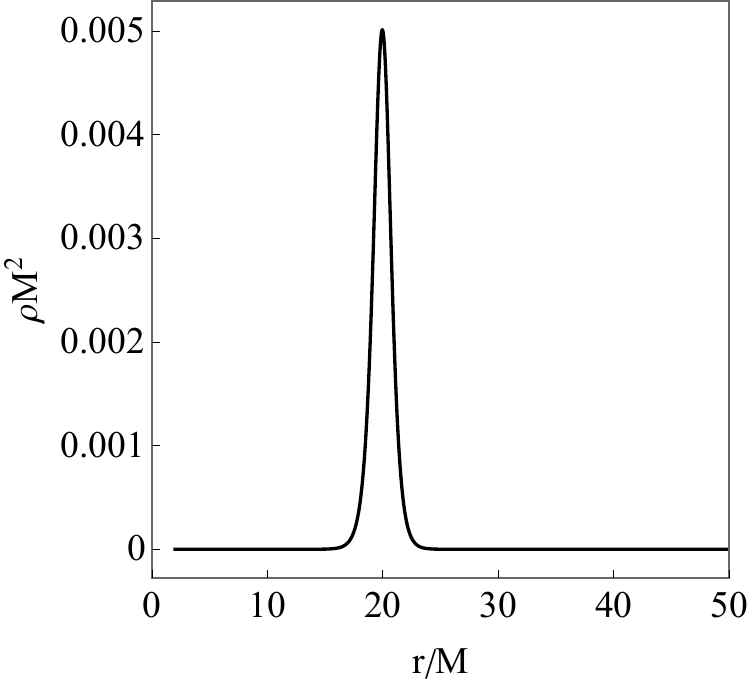}
\caption{The local energy density $\rho$ in Eq.~\eqref{LocalEnergyDensity} with $\epsilon_m=M$ and $a_m=20M$.}
\label{EnergyDistribution}
\end{figure}

Now, we assume that the local energy density is given by
\begin{equation}
\label{LocalEnergyDensity}
\rho= \frac{2\epsilon_m}{Mr^2}\cosh^{-2}\left(\frac{r-a_m}{M}\right),
\end{equation}
where $\epsilon_m$ and $a_m$ determine the peak value of the energy density and a location of the peak, respectively; $M$ is a length scale to be interpreted as a mass of a black hole below. The distribution is depicted in Fig.~\ref{EnergyDistribution}. From the $(t,t)$-component of the Einstein equation, the function~$m$ in Eq.~\eqref{sphericallysymspacetime} takes the form,
\begin{equation}
\label{massfunction}
m=M+\epsilon_m\left[1+\tanh\left(\frac{r-a_m}{M}\right)\right].
\end{equation}
Equations~\eqref{Prhorelation} and~\eqref{constraint} then determine the pressure~$P_t$ and the metric component~$A$.
\begin{figure}[htbp]
\centering
\includegraphics[scale=0.68]{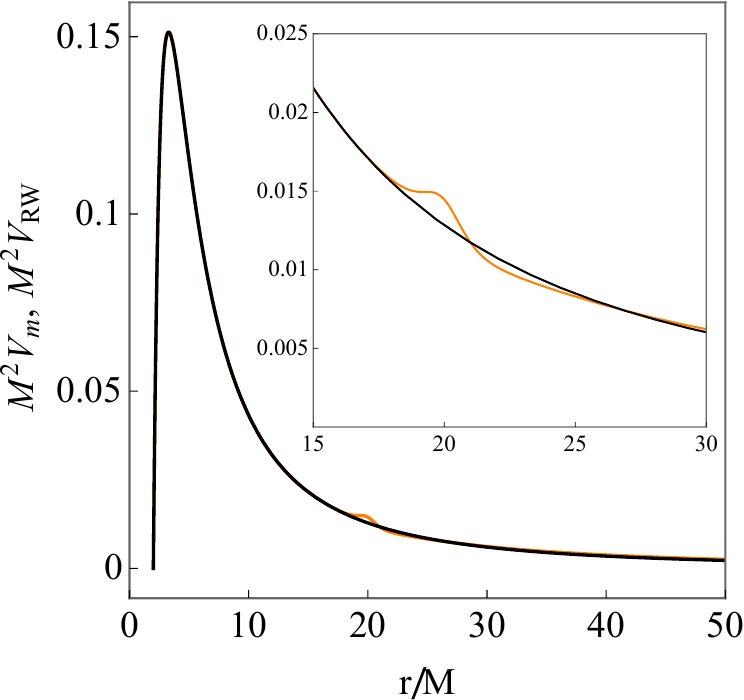}
\caption{For the quadrupolar mode~$\ell=2$, the Regge-Wheeler potential,~$V_{\rm RW}=(1-2M/r)[\ell(\ell+1)/r^2-6M/r^3]$~(black line) and the potential~$V_m$ in Eq.~\eqref{effectivepotentialVm}~(orange line) whose parameter set is the same as that in Fig.~\ref{EnergyDistribution}. The inset shows the enlargement around the peak of the energy density.}
\label{DeformedPotential_3}
\end{figure}

A linear gravitational perturbation to the geometry constructed above is written as
\begin{equation}
g_{\mu\nu}=g_{\mu\nu}^{(0)}+h_{\mu\nu},~~T_{\mu\nu}=T_{\mu\nu}^{(0)}+\delta T_{\mu\nu},
\end{equation}
where the superscript~$(0)$ denotes a background tensor field. As reviewed in Sec.~\ref{Section:tidalLoveandDissipationnumbers}, a symmetric tensor-field perturbation can be decomposed into two independent components, i.e., the even- and odd-parity perturbations, because of the parity invariance of the background. The linearized Einstein equation, $\delta G_{\mu\nu}=8\pi \delta T_{\mu\nu}$, leads to a radial equation with the harmonic decomposition, thereby obtaining the equation for the odd-parity perturbation~$\Phi$:
\begin{equation}
\frac{d^2\Phi}{dR^2}+\left(\omega^2-V_m\right)\Phi=0,
\end{equation}
with
\begin{equation}
\label{effectivepotentialVm}
V_m=A\left[\frac{\ell\left(\ell+1\right)}{r^2}-\frac{6m}{r^3}+\frac{1}{r^2}\frac{dm}{dr}\right],
\end{equation}
where $dR/dr=[A(1-2m/r)]^{1/2}$. The explicit form of $V_m$ is presented in Fig.~\ref{DeformedPotential_3}.

\section{Stability of the tidal response of a horizonless compact object}
\label{Appendix:HCOcase}

\begin{figure}[htbp]
\centering
\includegraphics[scale=0.7]{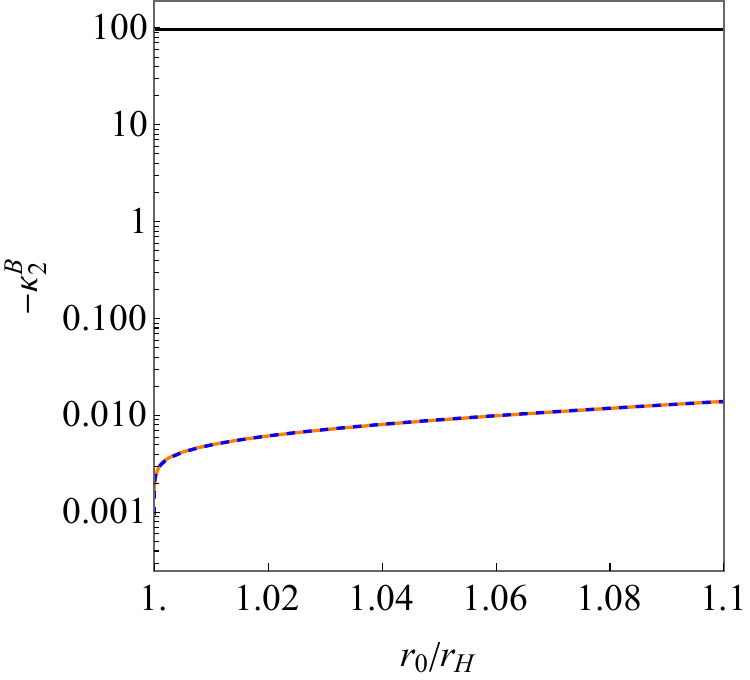}
\caption{The dependence of the quadrupolar magnetic-type tidal Love number with the opposite sign, i.e.,~$-\kappa_2^B$, on the radius at which the reflective boundary condition~\eqref{DBC} is imposed. The orange solid and blue dashed lines correspond to the results without the Gaussian bump. The former is the numerical result, while the latter is a model function,~$k_2^B=-0.0177/(1.0375+\ln \xi)$, whose logarithmic dependence on $\xi$ is consistent with the analytical result in Ref.~\cite{Cardoso:2017cfl}. The black solid line is the result with the Gaussian bump~\eqref{Gaussianbump} whose parameter set is the same as those in Fig.~\ref{ReResponseFunction_Gaussian1}. The black solid line is almost constant, $\kappa_2^B\simeq-96$, which is the almost same as the Love number of the Schwarzschild black hole with the same Gaussian bump~(see the orange line in Fig.~\ref{ReResponseFunction_Gaussian1}).}
\label{ReFofECO}
\end{figure}

We discuss stability of tidal Love and dissipation numbers when imposing a reflective boundary condition, i.e.,  the Dirichlet boundary condition,
\begin{equation}
\label{DBC}
\frac{A_{{\rm out},N}}{A_{{\rm in}, N}}=-\xi^{-2i\omega r_H},
\end{equation}
on Eq.~\eqref{GeneralPhiN}, at a radius~$r_0:=r_H(1+\xi)~(0<\xi \ll 1)$. Figure~\ref{ReFofECO} shows that in the absence of the Gaussian bump~(orange solid and blue dashed lines), the Dirichlet boundary condition leads to a non-zero tidal Love number whose absolute value increases for a larger boundary radius. The orange solid line and the blue dashed line are, respectively, the numerical result and a model function,~$k_2^B=-0.0177/(1.0375+\ln \xi)$. The results of the logarithmic dependence on $\xi$ and the negative Love numbers are consistent with the analytical result in Ref.~\cite{Cardoso:2017cfl}. We have checked that the dissipation numbers are quite smaller values than unity with or without the Gaussian small bump. 

In the presence of the Gaussian bump~(black solid line), the quadrupolar magnetic-type tidal Love number~$\kappa_2^B$ takes approximately $-96$. This value is less sensitive to the boundary radii and close to the tidal Love number of a Schwarzschild black hole with the same Gaussian bump~(see the orange line in Fig.~\ref{ReResponseFunction_Gaussian1}), i.e., $\kappa_2^B\simeq -96$. This implies that the property of the Gaussian bump mostly determines the value of the Love numbers in the current system. On the other hand, the dissipation numbers may tell the difference between black holes and horizonless objects.

\section{Useful formulas for special functions}
\label{Appendix:UsefulFormulas}

The useful formulas used in the main text are summarized based on Ref.~\cite{NIST:DLMF} in the following.
For the Gaussian hypergeometric functions,
\begin{equation}
\begin{split}
\label{Formula1for2F1}
~_2F_1\left(a,b;c;s\right)=&\frac{\Gamma\left(c\right)\Gamma\left(a+b-c\right)}{\Gamma\left(a\right)\Gamma\left(b\right)}\left(1-s\right)^{-a-b+c}\\
&\times \biggl[~_2F_1\left(-a+c,-b+c;-a-b+c+1;1-s\right) \\
&~~~~+\left(1-s\right)^{a+b-c}\frac{\Gamma\left(-a-b+c\right)\Gamma\left(a\right)\Gamma\left(b\right)}{\Gamma\left(a+b-c\right)\Gamma\left(-b+c\right)\Gamma\left(-a+c\right)}
\\
&~~~~~~ \times {}_2F_1\left(a,b;a+b-c+1;1-s\right)\biggr],
\end{split}
\end{equation}
and
\begin{equation}
\label{Formula2for2F1}
_2F_1\left(~,~;~;1-s\right)\big|_{s\gg1}=1+\mathcal{O}(1/|s|).
\end{equation}
For the confluent hypergeometric functions,
\begin{equation}
\label{Formula1forM}
M\left(a,b,s\right)\big|_{s\gg 1}=\frac{\Gamma(b)}{\Gamma(a)}e^ss^{a-b}\left[1+\mathcal{O}(1/|s|)\right]+\frac{\Gamma(b)}{\Gamma(b-a)}(-s)^{-a}\left[1+\mathcal{O}(1/|s|)\right],
\end{equation}
and
\begin{equation}
\label{Formula1forU}
U\left(a,b,s\right)\big|_{s\gg1}=s^{-a}\left[1+\mathcal{O}(1/|s|)\right].
\end{equation}
For the gamma function,
\begin{equation}
\label{FormulaforGamma1}
\frac{\Gamma\left(-2\ell-1\right)}{\Gamma\left(-\ell+2\right)}=\left(-1\right)^{\ell+1}\frac{\Gamma\left(\ell-1\right)}{2\Gamma\left(2\ell+2\right)},
\end{equation}
and
\begin{equation}
\label{FormulaforGamma2}
\frac{\Gamma\left(s\right)}{\Gamma\left(2s\right)}=\frac{\pi^{1/2}2^{-2s+1}}{\Gamma\left(s+1/2\right)}.
\end{equation}

\bibliographystyle{unsrt}
\bibliography{refs}

\end{document}